\documentclass[12pt]{spieman}  % 12pt font required by SPIE;
\usepackage{amsmath,amsfonts,amssymb}
\usepackage{graphicx}
\usepackage{setspace}
\usepackage{tocloft}
\usepackage{subcaption}
\usepackage{lineno}
\usepackage{siunitx}
\usepackage{lineno}
\title{\textit{S}pectro-polarimetry of \textit{HA}bitable \textit{P}lanet \textit{E}arth (SHAPE) on Chandrayaan-3: Instrument characteristics, calibration and onboard performance}

\author[a,b*]{Bhavesh Jaiswal}
\author[a]{Swapnil Singh}
\author[a]{Ravishankar Bangalore Thammannagowda}
\author[c]{Anand Jain}
\author[c]{Smrati Verma}
\author[a]{Reenu Palawat}
\author[a]{Brajpal Singh}
\author[d]{Bijoy Raha}
\author[d]{Sathyanarayana Raju}
\author[d]{Bhavesh Mendhekar}
\author[d]{Srinivasa Rao Kondapi}
\author[c]{Priyanka Das}
\author[c]{Rahul Waghmare}
\author[c]{Supratik Bose}
\author[c]{Supriya Verma}
\author[c]{Yogesh Prasad K R}
\author[c]{Praloy Karmakar}
\author[c]{Abhishek Kumar Singh}
\author[e]{Honey Gupta}
\author[c]{Balaraman Prabakaran}
\author[c]{Motamarri Srikanth}
\author[a]{Sankarasubramanian Kasiviswanathan}
\author[a]{Anuj Nandi}

\affil[a]{Space Astronomy Group, ISITE-Campus, U. R. Rao Satellite Centre, Outer Ring Road, Marathahalli, Bangaluru, 560037, India.}
\affil[b]{Indian Institute of Science, Bengaluru, 560012, India.}
\affil[c]{U. R. Rao Satellite Centre, Old Airport Road, Vimanapura post, Bangaluru, 560017, India.}
\affil[d]{Laboratory for Electro-Optics Systems, Bengaluru, 560058, India.}
\affil[e]{ISRO Telemetry, Tracking and Command Network, Bengaluru, 560058, India.}

\cftpagenumbersoff{figure}
\cftpagenumbersoff{table} 
\begin{document}
%\linenumbers
\maketitle

\begin{abstract}
The orbiter of the Chandrayaan-3 mission of the Indian Space Research Organisation (ISRO) carries an experimental payload called SHAPE (Spectro-polarimetry of HAbitable Planet Earth). This payload makes disc-integrated observations of Earth as an exoplanet, from the Moon as well as from the high altitude Earth orbit. The instrument consists of an Acousto-Optic Tunable Filter (AOTF) based near-infrared spectro-polarimeter making the measurements in the two orthogonally polarized directions using a pair of Indium-Gallium-Arsenide (InGaAs) detectors. The laboratory characterization of the flight model of the instrument included a relative measurement of the response of the two channels of the instrument. The field measurements of the instrument revealed a non-uniformity in the field response. In the light of non-identical response of the two channels and the non-uniformity of the field response, a theoretical model of such a polarimeter is developed to gain insights into the performance of a polarimeter. Analysis of lunar observations obtained within the SHAPE field of view indicates that the transmission ratio between the two polarized channels lies in the range 0.8–1.2. Such deviations from unity can introduce offsets in the measured polarization of up to $\sim10\%$. The suitability of SHAPE for studying the band polarization, defined as, the relative polarization within a spectral absorption band, is studied. Using the Moon observations and the theoretical instrumental model, the suitability of measuring the relative band polarization is demonstrated. A study of systematic biases in the band polarization demonstrate a maximum offset of less than $<1\%$ in the measured value of the band polarization. The initial results of Earth observations, spectra and flux measurements across phase angles, and the methodology for retrieving the band polarization from these observations are also discussed.
\end{abstract}

% Include a list of up to six keywords after the abstract
\keywords{Acousto-Optic, Spectro-polarimetry, Planetary atmosphere, Exoplanet, Chandrayaan-3}

% Include email contact information for corresponding author
{\noindent \footnotesize\textbf{*}First author email,  \linkable{bhavesh@ursc.gov.in} }

\begin{spacing}{1}   % use double spacing for rest of manuscript

\section{INTRODUCTION}\label{introduction}
Among all the planets in the solar system, Earth is unique in several ways. Firstly, because of the presence of liquid water on it and, secondly (perhaps more importantly) because of the evolution of life on it. Since the discovery of the first exoplanet around a Sun-like star \cite{1995Natur.378..355M}, questions about the presence of Earth-like planets around other stars have taken the center stage. Many of these questions will be confronted with the observations when we peek into the habitable zones of nearby stars using some of the most advanced telescopes proposed for the coming decades\cite{2022A&A...664A..21Q}. Such developments have led to increased curiosity over the disc-integrated spectral characteristics of a planet, which can help us identify the atmospheric and surface features. The Earth's disc-integrated signatures serve as a crucial reference for this study.

So far, the disc-integrated observations of Earth\cite{2010edpr.book.....V} have been limited to a few spacecraft observations which have flown at large inter-planetary distances like \textit{Galileo} \cite{1993Natur.365..715S}, \textit{EPOXI} \cite{2009ApJ...700..915C}, \textit{DSCOVR} \cite{2018RemS...10..254Y}, etc. Some of the limited observational attempts have proved to add interesting perspectives to the  existing knowledge about Earth \cite{2009ApJ...700..915C,2003JGRD..108.4710P,2006ApJ...644..551T, 2012Natur.483...64S}. Some of the modeling efforts \cite{2006AsBio...6...34T,2006AsBio...6..881T} further emphasize not only the importance of these observations but also the complexity involved in the retrievals of Earth-like planet. There have been several modeling efforts towards predicting spectro-polarimetric signatures \cite{2008A&A...482..989S,2012A&A...548A..90K, refId1, 2025A&A...697A.170R} -- many of which lack a direct confirmation as yet. This has led to the need for a dedicated experiment to observe Earth \textit{as an exoplanet}, i.e., disc-integrated observations of Earth from a large distance. For example project \textit{LOUPE} \cite{2012P&SS...74..202K,2021RSPTA.37990577K} and \textit{Earthshine} \cite{2022JATIS...8a4003B} have been proposed to observe Earth directly from the Moon.

Due to the lack of direct measurements of disc-integrated Earth, researchers have relied on ground-based observations of \lq Earthshine\rq\space in the visible and near-infrared (NIR) bands\cite{2003JGRD..108.4710P, 2006ApJ...644..551T, 2012Natur.483...64S, 2025A&A...702A.262R}. Earthshine is the reflected light from the day side disc of the Earth which is again reflected by the night side of the Moon and captured by the ground based telescopes, hence allows for the \textit{indirect} measurements of the disc-integrated light. The Earthshine measurements can cover a large range of phase angles (usually $\sim$ 40$^\circ$ to 140$^\circ$) as the Moon revolves around the Earth, and can also be equipped with polarization measurements. However, these observations suffer from depolarization due to the Lunar surface. \cite{1968JGR....73..649B}.

To bridge this gap in the observations and, in order to accomplish the dedicated observations of Earth as an exoplanet, we developed an \textit{experimental} payload, called SHAPE  \cite{2024arXiv241207416N} (acronym for Spectro-polarimetry of HAbitable Planet Earth) onboard the Chandrayaan-3 mission of the Indian Space Research Organisation (ISRO). The Chandrayaan-3 mission consisted of two main components: a lander module named \textit{Vikram} and a \textit{Propulsion Module}; the latter is hereafter referred to as the orbiter. The orbiter had very limited spacecraft capabilities — including no data storage system, minimal fuel and restricted maneuvering ability, and, carried a single payload, SHAPE. SHAPE uses an AOTF based optical filter and two InGaAs detectors to accomplish spectro-polarimetric observations in the wavelength range of 1000-1700 nm. The polarimetric performance of the experiment relied on the measurement of horizontally and vertically polarized beams (the two output beams of AOTF) by the two detectors. SHAPE has made several observations of Earth from the Moon. 

The experimental nature of the SHAPE payload is mainly because of two limitations in its design and operations, which are: [1] a non-uniform field response across the field-of-view (FOV) of the instrument and [2]
significant drift of the spacecraft during the observations.
The purpose of this paper is twofold: to highlight the field-dependent systematics in the instrument and to demonstrate the scientific results that can still be achieved with SHAPE despite these challenges. Motivated by the exciting scientific opportunities enabled by SHAPE, we systematically examine its limitations, quantify the measurements that remain feasible, and outline the methodology used to derive them. It is to be noted that after an initial two-month period in the lunar orbit, the spacecraft was maneuvered into a highly elliptical Earth-bound orbit, which provided new opportunities for lunar observations. This orbit allowed SHAPE to observe both Moon and Earth as integrated discs. In particular, the observations of the Moon enabled by the (high-altitude) Earth-bound orbit have been invaluable in assessing and quantifying SHAPE’s spectro-polarization performance. 

The investigations presented in this paper evolve in the following manner. First, we discuss the overall measurement methodology and operations in Section \ref{sec:config}. We then discuss the laboratory performance of the instrument, focusing on the methodology of dark measurements and the observed field response across its full FOV in Section \ref{sec:ins_char}. Although the laboratory measurements of the field response are not used to correct the onboard observations, its characteristics motivate a detailed examination of the polarimetric performance of an instrument with unequal throughput in the horizontally and vertically polarized beams. A theoretical model of such a polarimeter is analyzed in Section \ref{sec:ins_pol_model}, and its limitations in measuring absolute polarization, as well as its capability to measure the relative band polarization, are highlighted. Using the same model, we then analyze the Moon observations of SHAPE and quantify the  relative throughput of the two beams in Section \ref{sec:onb_perf}. Here, we examine the effect of spacecraft drift on the data, and establish the spectrally flat polarization of the Moon as observed by SHAPE. This analysis also allows us to estimate the systematic biases which can arise in the measurement of polarization. In the end, we show, in Section \ref{sec:Earth_Obs}, some of the Earth spectra observed by SHAPE across a large range of phase angles, consistently identifying the major gases in the atmosphere. We next assess the flux variations observed in SHAPE Earth observations and discuss the cautions required in using SHAPE data. Finally, we conclude with a discussion in Section \ref{sec:Disc}.

\begin{figure}[h!]
\begin{center}
\includegraphics[scale=0.8]{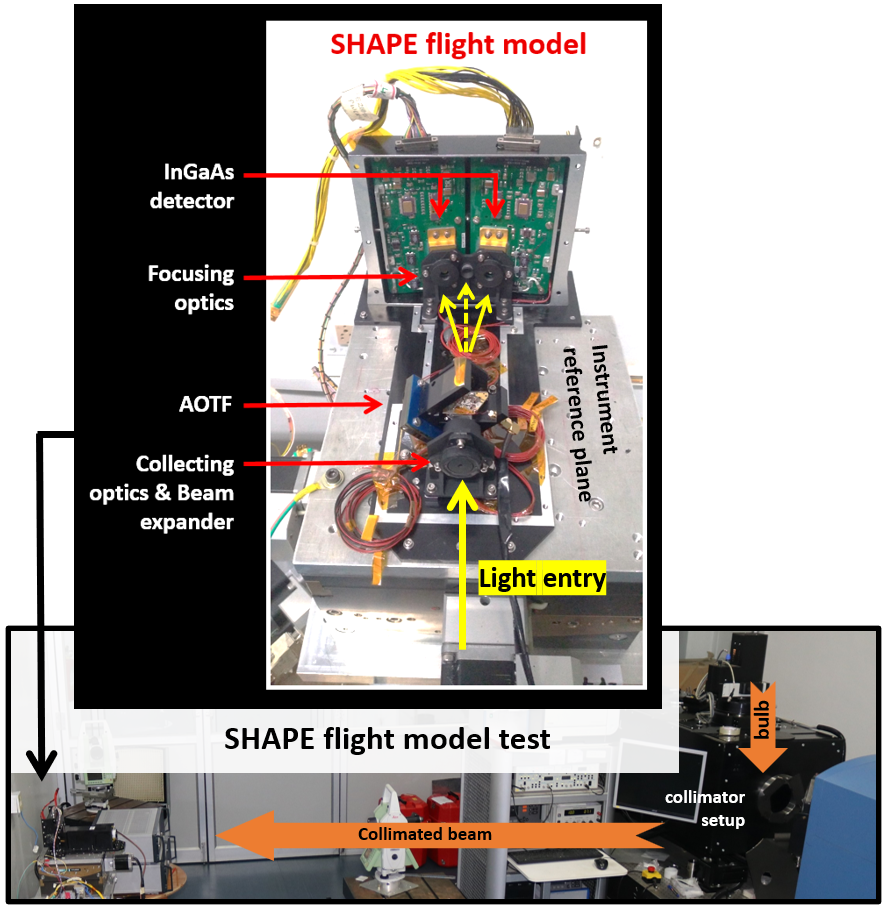}
\caption{SHAPE flight model and calibration setup. Various optics along with AOTF and detector are marked for the SHAPE flight model EODS package. The path of the light is marked with yellow arrows with the un-diffracted central beam, which is blocked, marked with dashed lines having the vertically and horizontally polarized beams on either side of it. The base plane of SHAPE is also the instrument reference plane. SHAPE flight model test setup is also shown in the bottom.}\label{fig_SHAPE_config}
\end{center}
\end{figure}

\begin{figure}[h!]
\begin{center}
\includegraphics[trim={0cm 4cm 0cm 3.5cm}, clip=true, scale=0.6]{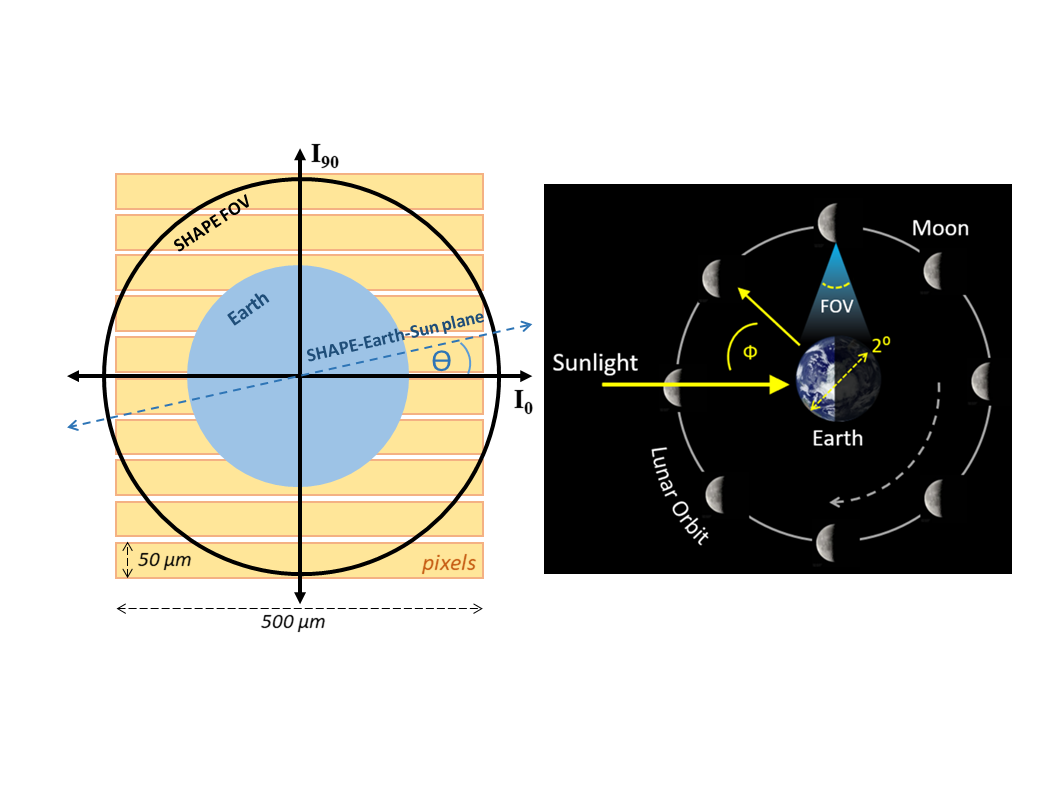}
\caption{[Left]The direction of polarization of the two detectors, $I_0$ $\&$ $I_{90}$, and the relative orientation of scattering plane (SHAPE-Earth-Sun plane for the case of Earth observations) is shown along with the relative orientation of the pixels in the background. [Right] SHAPE observation geometry for observing Earth from the Moon.}\label{fig_SHAPE_axes}
\end{center}
\end{figure}

% \begin{figure}[h!]
% \begin{center}
% \includegraphics[scale=0.6]{pol_modulation.png}
% \caption{Polarization modulation of the two beams as recorded in the flight model.}\label{fig_pol_modul}
% \end{center}
% \end{figure}

\section{SHAPE: METHODOLOGY \& OPERATIONS}\label{sec:config}
A detailed description of the laboratory model of Acousto-Optic Tunable Filter (AOTF) and the concept of the instrument along with the design optimization, is elaborated in Refs. \citenum{2015ExA....39..445A, 2022JATIS...8d4007J}. The details of the SHAPE configuration and operations can also be found in Ref. \citenum{2024arXiv241207416N} but for completeness we cover some essential aspects of SHAPE configuration here too.

SHAPE is a lightweight compact instrument which is designed to operate from the orbiter of Chandrayaan-3 spacecraft. The subtended angle of Earth is about \ang{2} at the Moon. The input optical aperture of the instrument is 2 mm, after which the beam inside the instrument goes to the AOTF which divides the beam into two orthogonal polarizations. As shown in Figure \ref{fig_SHAPE_config}, these beams are then collected by two focusing lenses and are focused onto two individual InGaAs pixelated (linear array) detectors. The wavelength tunability of the AOTF using the external radio frequency (RF) source allows spectral scanning across the wavelength range of SHAPE.  A previous example of a similar AOTF based instrument used for polarimetric studies is found in SPICAV (on-board Venus express), which was used to study the scattered polarization from Venusian clouds \cite{2015P&SS..113..159R}. Figure \ref{fig_SHAPE_config} shows the instrument configuration of SHAPE and the observation geometry is shown in Figure \ref{fig_SHAPE_axes}. The instrument has three packages: EODS (Electro-Optic Detector System; includes optics and detectors), EODS-Electronics and RFS (Radio Frequency Source; includes RF synthesizer, driver and power amplifiers). Here, we describe the measurement methodology of SHAPE.

For the measurement of linear polarization, the scattering plane (i.e., the plane containing the incident and scattered light directions) is considered to be the reference plane, which in this case is also the Sun-Earth-Moon plane. Using the intensities measured in the direction parallel ($I_0$) and perpendicular ($I_{90}$) to this plane, the Stokes $I$ and $Q$, in the plane of scattering, are defined in the following manner\cite{1974SSRv...16..527H}, as:

\begin{equation}\label{eq_signal1}
{I} =I_{0^\circ}+I_{90^\circ} \quad \mathrm{\&} \quad {Q} = I_{0^\circ}-I_{90^\circ}
\end{equation}
and
\begin{equation}\label{eq_DOLP1}
-\frac{Q}{I} = \frac{I_{90^\circ} - I_{0^\circ}}{I_{90^\circ} + I_{0^\circ}}.
\end{equation}

We define Eq. \ref{eq_DOLP1} as the degree of linear polarization ($DOLP$). The Stokes $U$ and $V$ are usually three to four orders of magnitude smaller than Stokes $Q$ when the measurements are made in the scattering plane and hence they are ignored and not included in the definition of $DOLP$ here (for example, see Refs. \citenum{2015P&SS..113..159R, 2022A&A...664A.172T}). If the measurements are made in a plane which is rotated by an angle $\theta$ with respect to the scattering plane, the Stokes vector gets multiplied with a rotation matrix and the new Stokes vector $I'$, $Q'$, $U'$ and $V'$ are obtained as: 

\begin{equation}
	\begin{bmatrix}
		1 & 0 & 0 & 0\\
		0 & \cos2\theta & \sin2\theta & 0\\
		0 & -\sin2\theta & \cos2\theta & 0\\
		0 & 0 & 0 & 1\\
	\end{bmatrix} \times \begin{bmatrix}
		I\\	Q\\	U\\	V\\
	\end{bmatrix}=\begin{bmatrix}
		I'\\ Q'\\ U'\\ V'
	\end{bmatrix}.
\end{equation}
where, essentially $I$ remains unaffected while $Q$ gets multiplied by $\cos 2\theta$. A relative orientation of the direction of polarization of the two beams and the scattering plane is shown in Figure \ref{fig_SHAPE_axes}.

In SHAPE, with respect to the base plane of the instrument (instrument reference plane), Detector-1 (measures $I_1$) measures the horizontal polarization and Detector-2 (measures $I_2$) measures the vertical polarization. Considering the transmission efficiency of the horizontal and vertical beams to be $\alpha_1 $ and $\alpha_2$ respectively, Equations \ref{eq_signal1} and \ref{eq_DOLP1} can be written as:

\begin{equation}\label{eq_signal1_FM}
{I} = \frac{I_1}{\alpha_1} + \frac{I_2}{\alpha_2} \quad \mathrm{and} \quad {Q}\cos2\theta = \frac{I_1}{\alpha_1} - \frac{I_2}{\alpha_2}
\end{equation}
where, $I_1$ and $I_2$ are measured intensities, and
\begin{equation}\label{eq_signal2_FM}
\frac{Q}{I}\cos2\theta = \frac{(\alpha_2/\alpha_1)I_1-I_2}{(\alpha_2/\alpha_1)I_1+I_2}
\end{equation}
The uncertainties in the measurements of $DOLP$ can be propagated from the uncertainties in the measurements of $I_1$ and $I_2$ and the details of this error propagation are given in Appendix \ref{appendix2_error}.

\begin{table}[h!]
	\caption{Major instrument specifications of SHAPE.}\label{tab_iconf}
	\begin{center}
	\begin{tabular}{ l  l }
		\hline
		Spectral range      & $1.0-1.7$ {\textmu}m \\
        RF sweep range for AOTF     & $80-135$ MHz  \\
        Spectral resolution & $2-4$ nm      \\\hline
		Input aperture      & 2 mm        \\\hline
		Integration time    & 10, 20, 50, 100, 200, 500, 1000 ms \\\hline
		Light polarization at output & Two orthogonal linear polarizations \\
                                        & Detector-1 (D1): Horizontal Polarization \\
                                        & Detector-2 (D2): Vertical Polarization \\\hline
	\end{tabular}
	\end{center}
\end{table}

\subsection{Instrument settings}

SHAPE uses two identical InGaAs detectors with associated electronics. Each detector has a one dimensional array of pixels, where each pixel is $500$ \textmu m in width and $50$ \textmu m in height (see Figure \ref{fig_SHAPE_axes}). The detector data are read out  and processed via the read-out and processing electronics. The detector voltage from each pixel is digitized with a 11 bit Analog to Digital Converter (ADC) and mapped onto a range of 0 to 2047 ADC channels, also called ADU (Analog-to-Digital Units). The optics design ensures that the focusing of the Earth-image on the pixels within a circular spot of $< 500$ \textmu m $\times 500$ \textmu m and hence the image area is not expected to increase beyond 10 pixels. The noise of the individual pixels is $< 2.6$ mV  or $< 2$ ADU.

There are several parameters in the onboard operation of SHAPE which can be customized based on the user requirements. The dynamic range of the optical signal can be accommodated by operating the instrument at a range of integration times ranging from 10 ms to 1000 ms along with the power of AOTF (which is proportional to the diffraction efficiency) from 0.5 to 2.0 Watt. The spectral observations in SHAPE are accomplished in the following manner: each wavelength of the spectrum is recorded for a finite time (set by integration time) and all the wavelengths in a spectrum are recorded in a sequential manner to complete one spectrum. SHAPE also has a control on the wavelength range which can be recorded (using RF start and end values) and also the frequency step-size at which it can be recorded. The wavelength steps are given by the steps in the RF \cite{2015ExA....39..445A,2022JATIS...8d4007J} and are independent of the spectral resolution of the instrument. Given the severe limitation of the onboard memory size of the experiment ($\sim$ 17 kbytes), the operations need to be optimized such that the maximum number of observations can be carried out before the memory is completely filled. Due to significant random drifts of the spacecraft boresight (reaching a maximum of up to $\sim\ang{2}$ per spectral scan), the spectral range of the operations was later restricted to approximately 1250 nm to 1700 nm to ensure that the most important portion of the spectrum — covering all major gases ($\mathrm{H_2O}$, $\mathrm{O_2}$ and $\mathrm{CO_2}$) — is captured more quickly (in half the time than full spectrum), thereby attempting to minimize the impact of spacecraft drift on consecutively recorded spectra.

\section{INSTRUMENT CHARACTERISTICS: Lab performance}\label{sec:ins_char}
Comprehensive ground calibration of the instrument, including radiometric, spectroscopic, polarimetric, and field calibration, has been carried out and incorporated into a ground calibration database. After the flight model of SHAPE was integrated with optics, detector electronics and RF unit, it was tested for its integrated performance, spectral and polarimetric response. This setup, shown in Figure \ref{fig_SHAPE_config}, was created at the Laboratory for Electro-Optic Systems (LEOS). The general idea behind all the tests is to characterize the broadband spectro-polarimetric performance for the entire FOV. Since Earth is an extended object as seen from Moon (with an angular extent of $\sim\ang{2}$), it is essential to study the full field performance of the instrument. Here, our emphasis is to discuss the instrument characteristics as observed in the laboratory which serve as a guide to understand the onboard performance. 

\subsection{Spectral Calibration} 
\begin{figure}[h!]
\begin{center}
\includegraphics[scale=0.35]{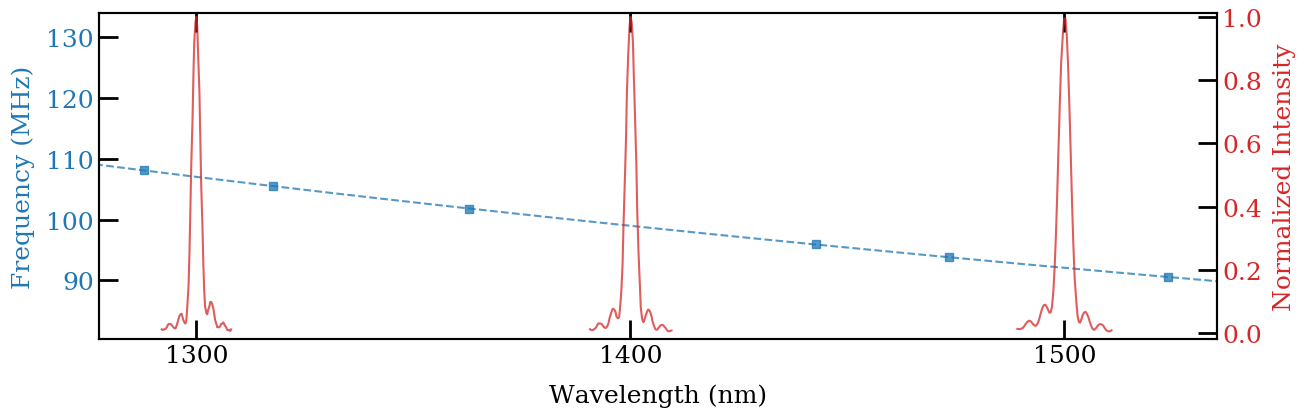}
\caption{RF$-\lambda$ relation for the output beam corresponding to horizontal polarization is shown in blue. The measured data points in the lab are represented by blue squares and best-fit function is given by the blue dashed line. The normalized $sinc^2$ functions at 1300 nm, 1400 nm and 1500 nm are shown using the solid red lines.}\label{fig_speccal}
\end{center}
\end{figure}

The details of spectral calibration of the laboratory model AOTFs are covered in Refs. \citenum{2015ExA....39..445A, 2022JATIS...8d4007J} and that of SHAPE flight model are covered in Nandi et al. 2026 (under review). However, here we cover the spectral calibration of SHAPE for completeness. In spectroscopic calibration, the spectral response, spectral resolution, and the frequency to wavelength tuning (RF–$\lambda$) relation of the instrument were characterized pre-launch following the methodology in Refs. \citenum{2015ExA....39..445A, 10.1117/1.JATIS.7.3.035001}. The RF$-\lambda$ relation was derived using standard emission lines from a Krypton lamp, while the spectral response of the AOTF ($sinc^2$) was estimated using monochromatic inputs generated through a broadband source coupled with a monochromator over the operational wavelength range. Figure \ref{fig_speccal} shows the RF$-\lambda$ relation obtained along with the spectral response functions at three wavelengths. The field measurements discussed in the subsequent section of this paper were also conducted at multiple wavelengths.

\subsection{Spectral and Field response} \label{sec:spec_field_response}
To study the full field response of the instrument, SHAPE was mounted on a movement stage which is capable of precise angular movement (with an error of  \ang{0.01}) in azimuth (horizontal) and elevation (vertical) angles. It is worth noting here that the detector pixels of the linear array are arranged in the vertical direction. For this study, we use a broadband source (a halogen bulb) and a collimator setup as shown in Figure \ref{fig_SHAPE_config}. A linear polarizer is mounted at the optical input of the SHAPE at \ang{45} with respect to the instrument base plate, such that equal intensities of the horizontally and vertically polarized light enters into the instrument. Essentially, this arrangement ensures that even though the light-source can be polarized, the intensities entering into SHAPE are equal, mimicking an unpolarized source. This setup allows us to create a broadband field response of SHAPE by doing a full spectral scan of AOTF at a large number of field angles.

%\begin{figure}[h!]
%\begin{center}
%\includegraphics[scale=0.4]{SHAPE_setup_LEOS.PNG}
%\caption{The laboratory setup used to study the response of the flight model of SHAPE. The collimated beam of light comes out from the collimator setup shown on the right. The beam then goes to the SHAPE flight model which is mounted on a rotation stage shown on the left of the figure. The SHAPE model is shown in the inset of the image.}\label{fig_SHAPE_setup}
%\end{center}
%\end{figure}

\begin{figure}[h!]
\begin{center}
\includegraphics[scale=0.55]{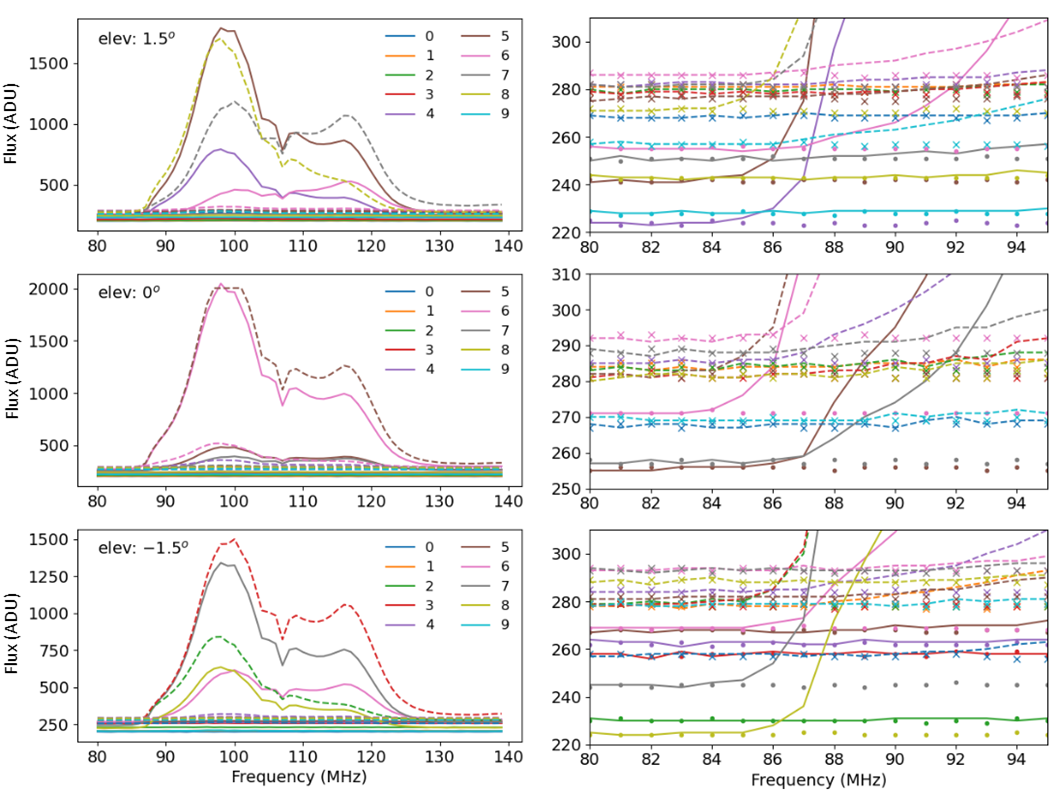}
\caption{Spectral scans recorded at different field angles. Top: Elevation \ang{1.5}, middle: Elevation \ang{0} and bottom: Elevation \ang{-1.5}. The azimuth angle is considered to be \ang{0} for all the cases. [Left panel] The spectral scan of a halogen bulb for the 10 pixels (shown with different colors) in the two detectors. A total of 10 pixels, as depicted in \ref{fig_SHAPE_axes}, for $D1$ and $D2$ are shown with solid and dashed lines respectively. [Right panel] The dark values recorded for the two detectors are shown using `dots' and `crosses' for $D1$ and $D2$ respectively using the same color scheme as left panel. For a comparison, the spectral scan in the illuminated condition (as shown in the left panel) is also Over-plotted with lines using the same color scheme and style as shown in the left panel. A range of only 80-95 MHz is chosen to clearly show that the flux in the illuminated condition (shown with lines) reaches the dark levels (shown with dots and crosses) for frequencies below 86 MHz.}\label{fig_signal_dark}
\end{center}
\end{figure}

The collimated light is incident on SHAPE aperture and the spectrum is scanned from 80 MHz (corresponds to $\sim 1700$ nm) to 135 MHz (corresponds to $\sim 1000$ nm). In the left panel of Figure \ref{fig_signal_dark} we show the full spectral scan of the halogen bulb observed at three different elevation angles and a fixed azimuth angle of \ang{0}. The scans are shown for 10 pixels of each detector (shown with different colors) where the signal is shown as ADU. As can be seen, the light is focused onto a few pixels and the rest of the pixels show very low counts (equal to the dark value). Since, for a collimated beam the spot is typically spread over three pixels, where one pixel is $50\mu m$ in height, it shows that the beam spot-size at the detector is $100\mu m-150\mu m$ in size. The drop in the intensity at higher RF frequencies (and lower wavelengths) is mainly due to the low efficiency of AOTF-RF system and the drop at the smaller RF frequencies (and larger wavelengths) is due to the drop in the quantum efficiency of the detector, as is also shown in Ref \citenum{2022JATIS...8d4007J}. Since the pixels in the detector are arranged in the vertical direction, any change in the field elevation angle manifests in the change in the location of the peak-signal on the pixels. As can be seen that the pixel with the highest counts shifts (from 5 to 6 to 7) as the elevation angle is changed (from \ang{-1.5} to \ang{0} to \ang{1.5}). It is noteworthy here that there is no shift in the peak-signal pixels with azimuthal angle because the pixels in the linear array are arranged in the vertical (elevation) direction. In the right panel of Figure \ref{fig_signal_dark} we show the `dark' measurement which were obtained by blocking the halogen bulb source and hence serve as a base-level. The signal in the `dark' measurement  of the two detectors consists of the base level of the detector-electronics along with any background scattered light. The `dark' spectra recorded at all the field angles for all the pixels do not show any spectral dependence. The observed signal at $\sim 80$ MHz (at the higher edge of the spectrum $\sim$1700 nm) in the `illuminated' conditions is seen to be dropping to a value which is equal to the `dark' value of that pixel. This shows that the instrument response is zero at $\sim$1700 nm wavelength, owing mainly to the zero quantum efficiency of the detector at these wavelengths. This also shows that 80 MHz RF frequency of AOTF also serves as a good reference value for the dark measurement.

\begin{figure}[h!]
\begin{center}
\includegraphics[scale=0.36]{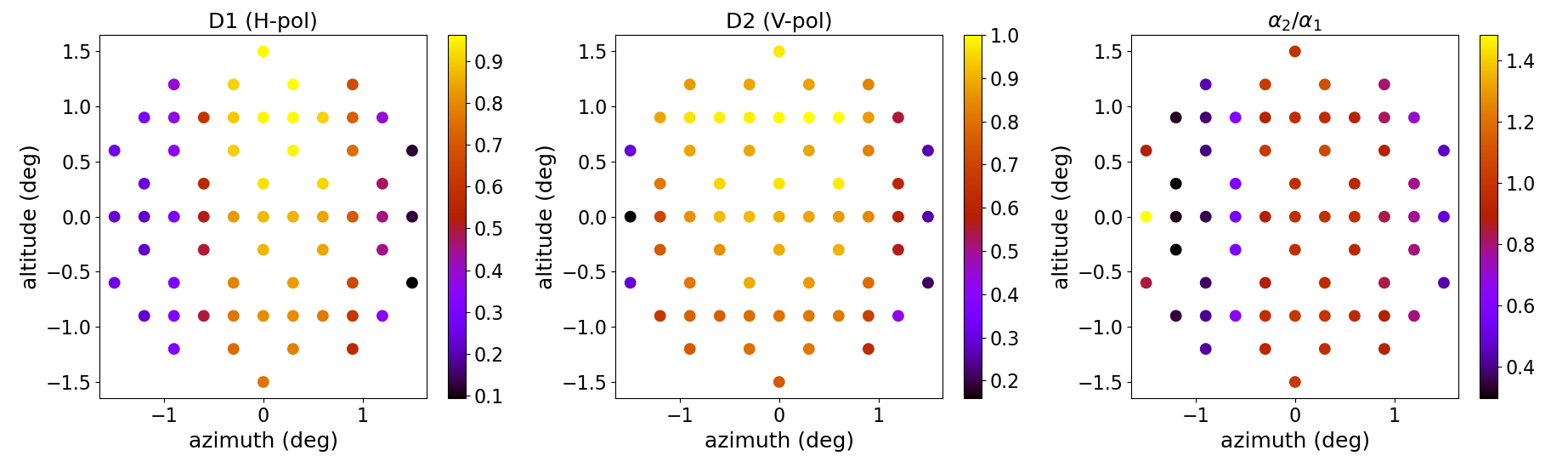}
\caption{Field measurements of SHAPE. Left and middle panels show the relative intensity as measured in different regions of the field for the two channels of SHAPE flight model. The values are normalized to the maximum intensity observed in detector 1 (D1). The right panel shows the $\alpha_2/\alpha_1$ as obtained from this field data.}\label{fig_field}
\end{center}
\end{figure}

To measure the full field response of the instrument we carried out measurements at a step of \ang{0.3} in the azimuth and elevation directions and then interpolate between them to create a circular field response of $\pm{\ang{1.5}}$. The response measured at 1500 nm wavelength is shown in Figure \ref{fig_field} for both the detectors. As can be observed, the response of the two fields is not the same for the two detectors and also has a non-uniformity in the field, especially in the azimuthal direction.  This non-uniformity has serious implications for the range of useful field which corresponds to meaningful flux levels when doing observations.

As per Eqs. \ref{eq_signal1_FM} and \ref{eq_signal2_FM}, to obtain the value of Stokes $Q$ and $I$, we need to know the transmission of the two beams, that is, the values of $\alpha_1$ and $\alpha_2$ individually. As discussed in Ref. \citenum{2022JATIS...8d4007J}, for such transmission measurements it is required to measure the light incident on the AOTF input aperture, which is done by removing the AOTF from the optical chain and measuring the un-diffracted beam by bringing the detector in the center of the optical chain. This flexibility is possible in the lab setup however is not available in the flight model. For this reason we rely on measuring the ratio of the transmission of the two beams, that is, $\alpha_2/\alpha_1$, which can be used, as per Eq. \ref{eq_signal2_FM}, to obtain the degree of polarization.

%4. The polarization modulation curve of the instrument was measured by using an un-polarized source and modulating it's plane of polarization with a rotating polarizer. The modulated light is measured in the two channels of the instrument. The modulation curve, as shown in figure \ref{fig_pol_modul}, is seen to follow a characteristic modulation of a linear polarizer -- a sinusoidal modulation.

As seen in Figure \ref{fig_field}, we find that field response of the two channels of the instrument is not uniform for the entire field, especially in the azimuthal direction. This kind of response can be caused due to vignetting in the optical path. This also leads to a non-identical field variation of the transmission in the two channels, especially in the azimuthal direction. It would lead to a variation in the value of $\alpha_2/\alpha_1$ (as shown in Figure \ref{fig_field}) throughout the field, mostly in the azimuthal direction. It is noteworthy here that $\alpha_2/\alpha_1$ includes the overall instrument response and not just the variation in the field.

\begin{figure}[h!]
\begin{center}
\includegraphics[scale=0.4]{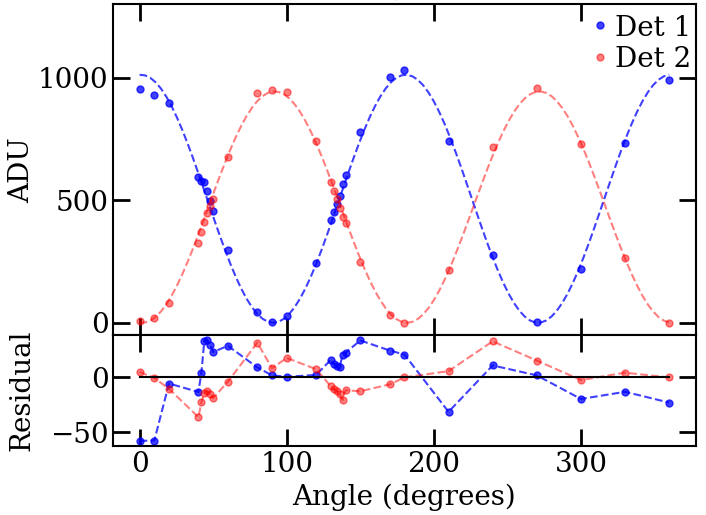}
\caption{Polarization modulation as measured by the flight model of SHAPE. The measurement was made for an input frequency of 101.8 MHz to the AOTF. The blue and red circles are the measured ADU values for Detector 1 and Detector 2 respectively. The dashed lines of the corresponding colours are the best fit function of the form $I= a  \times \mathrm{sin}^2(\theta – b)$. The bottom panel shows the residual values.}\label{fig_pol_modul_lab}
\end{center}
\end{figure}

As a standard test of polarization, we performed a polarization modulation test on SHAPE flight model. This was carried out by introducing a rotating polarizer between the instrument and an unpolarized source and recording the modulation in the two orthogonally polarized detector channels at various frequencies for the central field. One measurement at an AOTF RF frequency of 101.8 MHz is shown in Figure \ref{fig_pol_modul_lab}. The measured sinusoidal variation demonstrates the polarimetric capability of the flight model. Further details of the characterization and analysis are presented in Nandi et al. 2026 (under review). In the next section, we study a theoretical model to study the effect of retrieving the polarization from an instrument having a non-identical transmission in the two beams.

\section{INSTRUMENT CHARACTERISTICS: Instrumental Polarization Model}\label{sec:ins_pol_model}
As the non-uniformity of the field response and the unequal ratio of the horizontal and vertical channel is established now, we make an attempt to address the polarimetric performance of the instrument. The first step in that attempt is to theoretically study the polarimetric performance of a polarimeter with an unequal response in the horizontally and vertically polarized channel. 

\begin{figure}[!htbp]
\begin{center}
\includegraphics[scale=0.4]{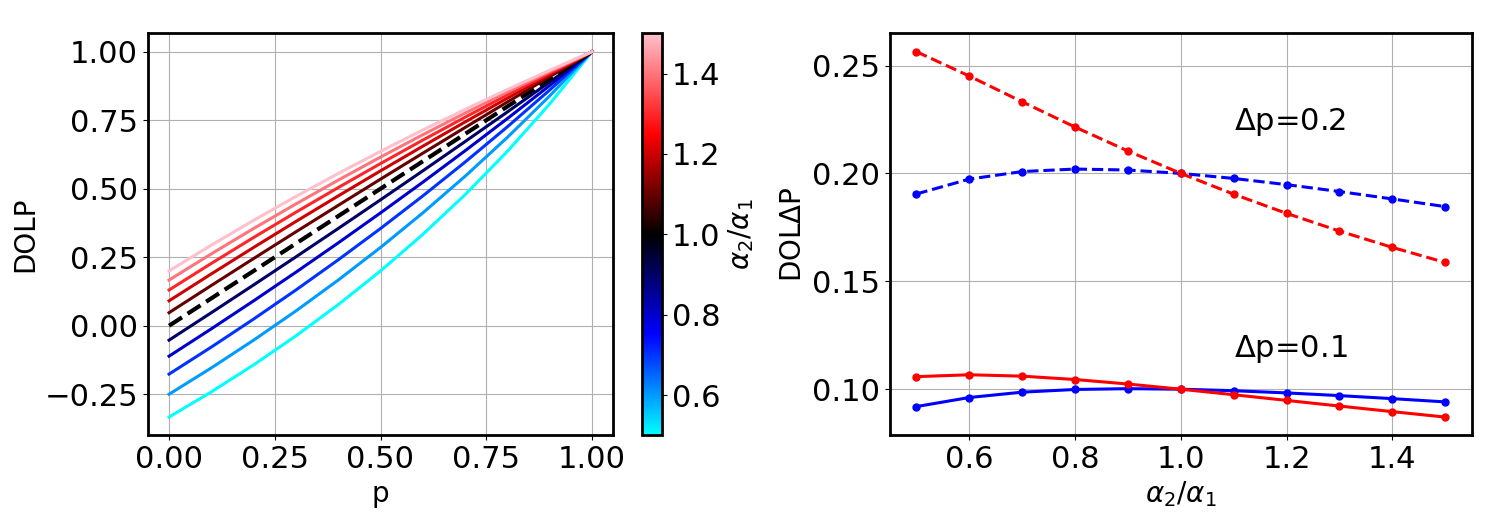}
\caption{[Left] The $DOLP$ calculated for a range of incident polarization $p$ as per Eq. \ref{eq_DOLP_estim} for a range of $\alpha_2/\alpha_1$ values from 0.5 to 1.5 in the steps of 0.1. The $DOLP$ for $\alpha_2/\alpha_1=1$ is plotted with a thick dashed line for reference. [Right] The $DOL\Delta P$ is plotted for two cases of $\Delta$p = 0.1 and 0.2 with solid and dashed lines respectively. For $\Delta$p=0.2 the cases with p1=0.4 and 0.2 are marked with red and blue lines respectively and for $\Delta$p=0.1 the cases with p1=0.3 and 0.1 are marked with red and blue lines respectively.}\label{fig_DOLP}
\end{center}
\end{figure}

It follows from Eq. \ref{eq_signal1_FM} and \ref{eq_signal2_FM} that if we use $I_1$ and $I_2$ to estimate the degree of linear polarization ($DOLP$), as per Eq. \ref{eq_DOLP1}, for a source having an inherent polarization $p$ ($=-Q\cos2\theta/I$), we get, 

\begin{equation}\label{eq_DOLP_alpha_frac}
DOLP = \frac{I_2-I_1}{I_2+I_1} = \frac{(\alpha_2/\alpha_1)(1+p)-(1-p)}{(\alpha_2/\alpha_1)(1+p)+(1-p)}.
\end{equation}
For an ideal polarizer, $\alpha_2/\alpha_1 = 1$ and $DOLP=p$. However, for  $\alpha_2/\alpha_1 \neq 1$, as in the case of SHAPE, the obtained $DOLP\neq p$. We do a theoretical experiment to find the $DOLP$ for a range of values of $\alpha_2/\alpha_1$ and $p$. In Figure \ref{fig_DOLP}, we plot $DOLP$ vs $p$ for a range of $\alpha_2/\alpha_1$ values and we see that for $\alpha_2/\alpha_1=1$ the system behaves like an ideal polarizer, where the $DOLP$ vs $p$ plot follows a straight line with slope=1. However for $\alpha_2/\alpha_1 \neq 1$ the relation starts to delineate showing that the $DOLP$ is no longer a representative of the incident polarization $p$ and it either over or under estimates $p$ depending on whether $\alpha_2/\alpha_1$ is larger or smaller than 1. Even for $p=0$ the $DOLP$ shows an offset (or a bias) in the measured polarization. With this exercise we understand that a system like SHAPE with $\alpha_2/\alpha_1\neq 1$ can result in a serious offset in the polarization depending upon the value of $\alpha_2/\alpha_1$ which, as can be seen from the left panel of Figure \ref{fig_DOLP}, can reach to $\sim100\%$ of the incident polarization. 

The above investigation reveals the limitation in SHAPE to study the continuum polarization of a source. However, Earth-like planets are expected not to just cause polarization in the continuum but also within the absorption bands of gases \cite{2014A&A...562L...5M}. For this reason, the relative polarization of absorption band to that of the continuum, called as band polarization or $\Delta$ polarization, can manifest in the SHAPE spectral-polarization observations. Hence, we next investigate the trend of $\Delta$ polarization, which we denote as DOL$\Delta$P (degree of linear delta polarization), which we define as the difference in $DOLP$ calculated at two values of $p$: $p1$ and $p2$ (for a fixed $\alpha_2/\alpha_1$):

\begin{equation}\label{eq_DOLP_estim2}
DOL\Delta P = DOLP (p1) - DOLP(p2).
\end{equation}

In a spectrum, $p1$ and $p2$ can correspond to the polarization at two different wavelengths, having an inherent spectral $\Delta$ polarization = $p1-p2$. It is noteworthy here that for a specific case of $\alpha_2/\alpha_1=1$, $DOL\Delta P$ = p1-p2 -- the inherent polarization of the source. In Figure \ref{fig_DOLP}, we plot the $DOL\Delta P$ vs $\alpha_2/\alpha_1$ for two cases of $\Delta p$ i.e. 0.1 and 0.2. For the case of $\Delta p$=0.1, two cases of $p1$ are considered i.e. 0.3 and 0.1 (corresponding $p2$ will be 0.2 and 0). For the case of $\Delta p$=0.2, two cases of $p1$ are considered i.e. 0.4 and 0.2 (corresponding $p2$ will be 0.2 and 0). From this exercise, we find that, for these cases, for a range of $\alpha_2/\alpha_1$ values centered within $0.8\lesssim \alpha_2/\alpha_1 \lesssim 1.2$ the obtained $DOL\Delta P$ using Eq. \ref{eq_DOLP_estim2} is close to the inherent $\Delta p$ of 0.2 and 0.1 within $10\%$ of its original value. 

This investigation reveals the retrievals of $\Delta$ polarization values which are close to the actual values, despite the serious offsets in the measurement of the absolute value of the polarization for a polarimeter having a non-identical response ($\alpha_2/\alpha_1 \neq 1$) in the horizontally and vertically polarized channels. In the next section, we discuss selected Moon observations to study the variation of $\alpha_2/\alpha_1$ in SHAPE field.

\section{ONBOARD PERFORMANCE EVALUATION WITH MOON OBSERVATIONS}\label{sec:onb_perf}
At the time of writing this paper, SHAPE has operated for about 3 years since its first operation. During the course of last two years of operation, the spacecraft has changed its orbit. From the first-light on 21 Aug, 2023 till 04 Oct, 2023, the spacecraft was operated from the Moon-bound orbit \cite{2024arXiv241207416N}. After 04 Oct, 2023 onward, the spacecraft is operated from a highly elliptical Earth-bound orbit which provides a unique opportunity of observing disc-integrated Moon for the purpose of calibration. The Earth bound orbit also allows us to do the Earth observations. All the three packages of SHAPE are mounted on the negative pitch panel of the spacecraft with SHAPE boresight in the positive roll direction of the spacecraft.

In the Moon-bound orbit, as the orbiter revolves around the Moon in an approximately 2 hour orbit, once in each orbit the spacecraft is visible to Earth for an approximate 1 hour period. During this period, based on the visibility of the spacecraft at the Bengaluru tracking station (ISTRAC), the observations of Earth are carried out. Each SHAPE operation, from switch ON to switch OFF, takes about an hour and involves re-orientation of the spacecraft (see Table \ref{table_event_schedule}). Since Chandrayaan-3 orbiter does not carry reaction wheels for a precise 3-axis stabilization of the pointing during the SHAPE observations, the SHAPE observations suffer from random drifts of the FOV during the observations, though it is largely ensured that Earth does not go completely out of the FOV (drift $< \ang{2}$) during the imaging operations. The coarse control of the SHAPE bore-sight is achieved by firing spacecraft thrusters. The knowledge of the FOV during the operations is available. An example of spacecraft FOV drift is shown with respect to the Earth observations in Figure \ref{fig_SHAPE_Earth_drift}.

Each observation sequence of SHAPE consists of re-orientation of the spacecraft from its current orientation to point towards Earth / Moon, including acquisition of dark. The spectral scanning (scientific observation) starts after acquiring the Earth/Moon in the SHAPE FOV. Depending upon the command sequence, the spectral scanning starts from start frequency to end frequency with a particular step size of the RF scan. At each frequency the signal is acquired on the detector for a fixed time given by the integration time of the detector. Depending upon the spectral range and the step size and integration time, one spectral scan can take anywhere from few 10s of milliseconds to several 10s of seconds. During the spectral scanning, the spacecraft undergoes several cycles of pointing evaluation and re-adjustment of the pointing to ensure the source (Earth/Moon) remains within the instrument FOV. This is primarily due to attitude control that is managed by thrusters -- which, after a firing sequence, also imparts the spacecraft with a residual momentum that leads to the continuous drift of the spacecraft during the spectral scanning. Several spectral scans, in a consecutive manner, are acquired during one observation. The total time taken for all the scans, depending upon the scanning configuration, can take from a few seconds to several minutes and during this time the spacecraft can undergo several drift cycles in an independent manner. The Table \ref{table_event_schedule} marks the sequence of events in a typical observation cycle of SHAPE.
 
\subsection{Detector reference levels: Dark counts}
The reference level of the detector is also referred to as the `dark' level. The dark level in the SHAPE observations is defined as the detector output in the absence of any AOTF-dispersed light in the direction of the detector. The dark level, in terms of detector output in ADU, mainly consists of the base level of the detector output (in no-light condition there is a finite output of the detector in ADU; called base level) and also the contribution of the scattered light falling onto the detectors. The scattered light can be caused by any photons which, after entering the optical aperture, can deviate from the optical path at any stage of the optical path and can get scattered within the optical assembly of the instrument and contribute to the detected light by falling onto the detectors. Essentially, the dark counts in the SHAPE detector are caused by the scattered light within the system on top of the base level of the detector. In all our laboratory tests, the base levels for the detectors have remained in the range of $200-210$ and $230-240$ bins of the ADC for all the pixels of Detector 1 and Detector 2 respectively. During our onboard measurements we do measure the dark counts, by switching off the AOTF, at the end of the spectral scanning. Since the slow drift of the spacecraft is seen to affect the observed light falling onto the detectors, it may also affect the dark measurements during the spectral scan. Hence, for a stronger constrain on the dark we rely on the dark measurements obtained during the spectral scans. As seen in the laboratory measurements in section \ref{sec:spec_field_response}, the spectral measurements obtained at $\sim$1700 nm ($\sim$80 MHz) can serve as an excellent proxy for the dark measurements. As we show in several Lunar spectral observations discussed in section \ref{sec:Observations_of_Moon_to_study_field}, the dark counts observed at the $\sim 1700$ nm wavelength ($\sim$80 MHz of RF), at the beginning of each spectrum, show a slow variation with time. The dark count for each spectral point can be estimated by interpolating between the dark observed at the beginning of each spectrum. A typical variation of dark in the case of Moon observations is shown in Figure \ref{fig:Moon_SHAPE_spectrum_dark}. As can be seen in the bottom panels of these figures, the dark counts have slowly varied by a maximum of $\sim2\%$ during the entire duration of the observations where 13 spectra are recorded and by $<1\%$ for the two consecutive spectra. Dark counts in the case of Earth observations also show similar variations.

\begin{table}[h!]
\centering
\begin{tabular}{|l|l|l|}
\hline
\textbf{Sl. no.} & \textbf{Event} & \textbf{Typical Time} \\
\hline
1. & Spacecraft Reorientation for & T0  \\
 & Observation &  \\
2. & SHAPE Switch ON & T1 = T0 + 15 mins  \\
3. & Dark-1 acquisition & T2 = T0 + 17 mins  \\
4. & SHAPE spectra acquisition start & T3= T0 + 19 mins  \\
5. & SHAPE spectra acquisition end & T4 = T3 + Variable  \\
6. & Dark-2 acquisition & T5 = T4 +2 secs  \\
7. & Spacecraft Reorientation for & T6 = T4+ 2 mins  \\
   & Data transmission  &  \\
8. & Data Transmitted & T7 = T4 + 11 mins  \\
9. & SHAPE Switch OFF & T8 = T4 + 35 mins \\
\hline
\end{tabular}
\caption{Representative event schedule outlining observation sequence and dark frame acquisition stages.}
\label{table_event_schedule}
\end{table}

\subsection{Observations of Moon}\label{sec:Moon_obs}

SHAPE, after being brought to the Earth-bound orbit, has performed several disc-integrated observations of Moon. It is noteworthy here that during the Moon observations the spectral range has been reduced to $\sim 1250$ - $ 1720$ nm and the spectral sampling is reduced to $\sim$ 15 nm, in order to reduce the spectral scan time which allows us to make consistent observations before spacecraft accumulates detrimental amount of drift. Due to low flux of the Moon, many of the observations are done at a maximum integration time of 1000 ms. These observations are scheduled as and when the apparent size of the Moon is sufficiently large ($\gtrsim \ang{0.5}$) in the SHAPE FOV to cause significant increase in flux. The Moon observations have been carried over a large range of lunar phase angles.

\begin{figure}[h!]
\begin{center}
\includegraphics[scale=0.35]{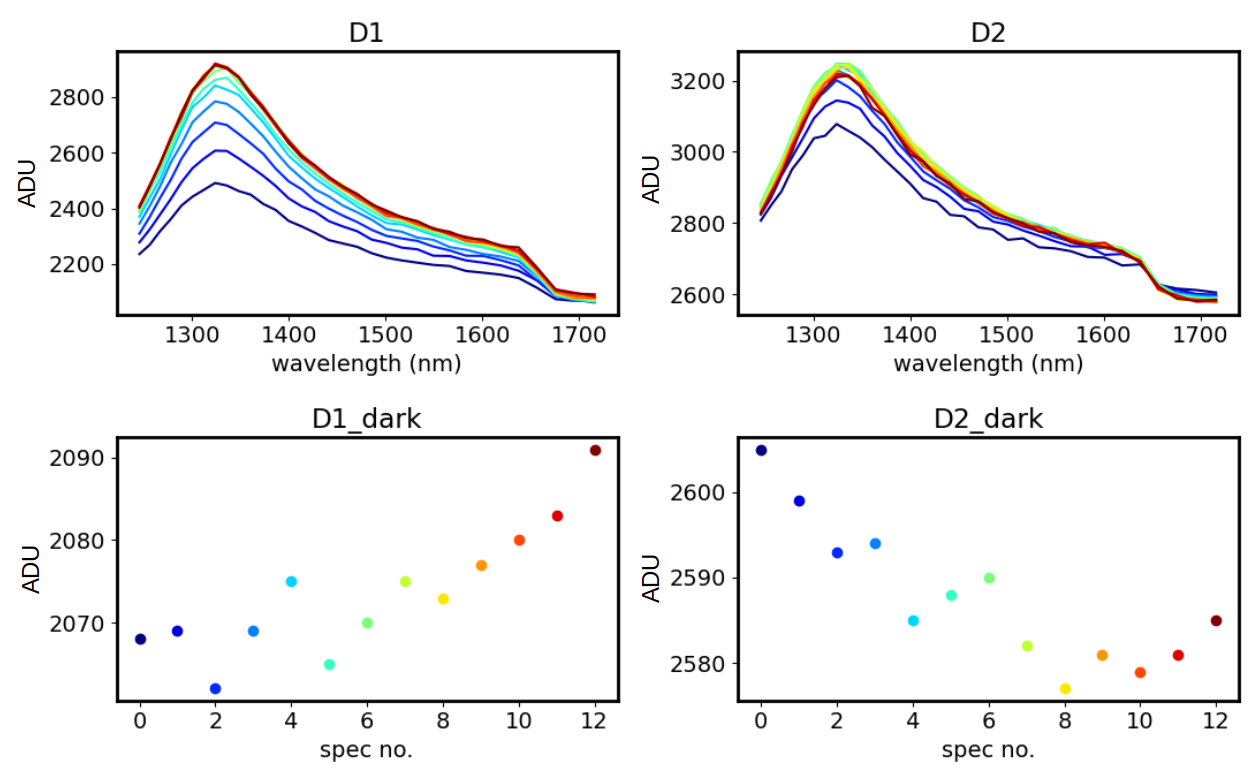}
\caption{Selected Moon spectra [top] and their corresponding detector dark values [bottom] observed in the two detectors (corresponding to the spectrum number given on the X-axis), for the 18 Mar 2024 observation. The raw spectra are shown without subtracting the dark. The details of this observation are given in Table \ref{table_moon_obs} and the movement of the Moon within the SHAPE FOV is shown in Figure \ref{fig_lunar_pol} for all these spectra.}\label{fig:Moon_SHAPE_spectrum_dark}
\end{center}
\end{figure}

\begin{figure}[h!]
\begin{center}
\includegraphics[scale=0.7]{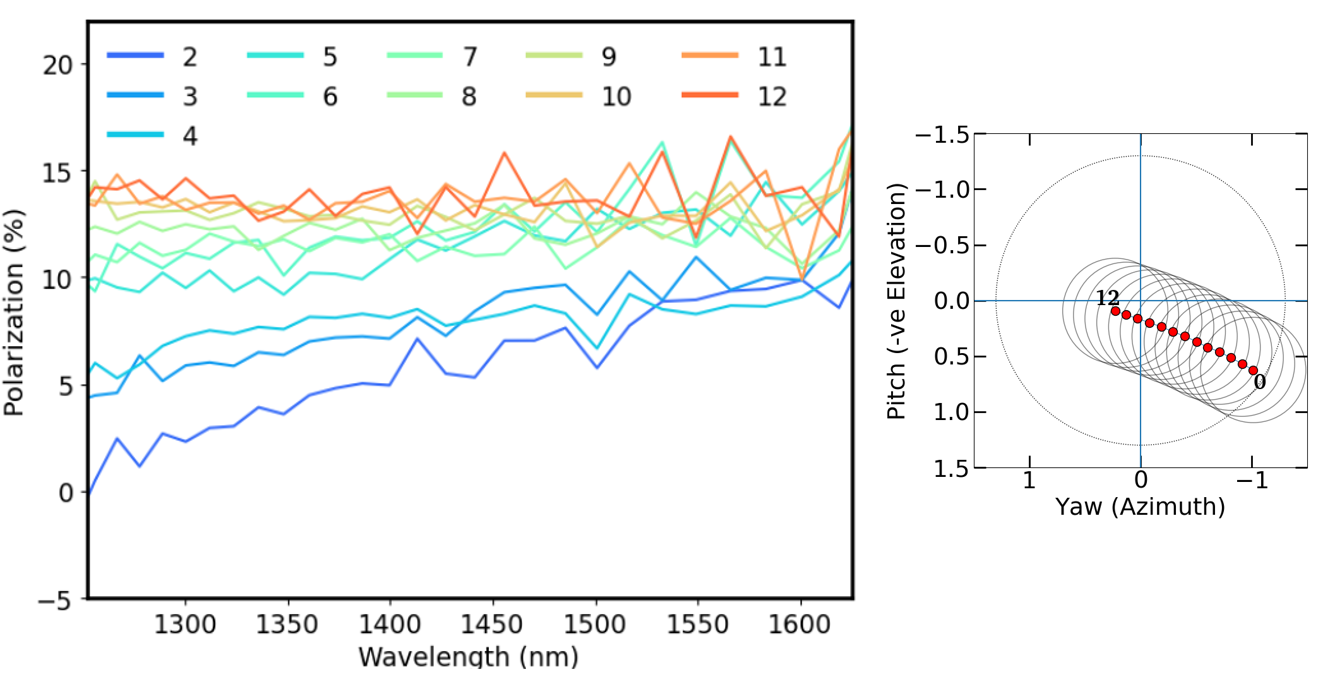}
\caption{Observed $DOLP$ (as $\%$ Polarization) [on left] for various Moon observations and the corresponding location of the Moon within the FOV for the 18 Mar 2024 observations of the Moon [on right]. As Moon, shown as black circles, swings across the SHAPE FOV (shown as large black circle), the spectrum as well as $DOLP$ shift accordingly. Observations for record no. 0 and 1 are not shown as they are outside of the FOV.}\label{fig_lunar_pol} %PLEASE USE Fig8_sub to replace the plot on bottom right.
\end{center}
\end{figure}

The observations of the Moon have allowed us to study the onboard instrument characteristics and more importantly to investigate the source of inconsistent observations and limitations they introduce. As mentioned earlier, the major source of systematic variation in the flux recorded is caused mainly due to the non-uniform field response and drift in the spacecraft pointing which builds up, in random directions, during the time of observations. We show one such observation, happened on 18 March 2024, which demonstrates these limitations. The recorded spectra are shown in Figure \ref{fig:Moon_SHAPE_spectrum_dark} and the corresponding locations of the Moon center in the SHAPE FOV are shown in Figure \ref{fig_lunar_pol}.

\begin{figure}[h!]
\begin{center}
\includegraphics[scale=0.7]{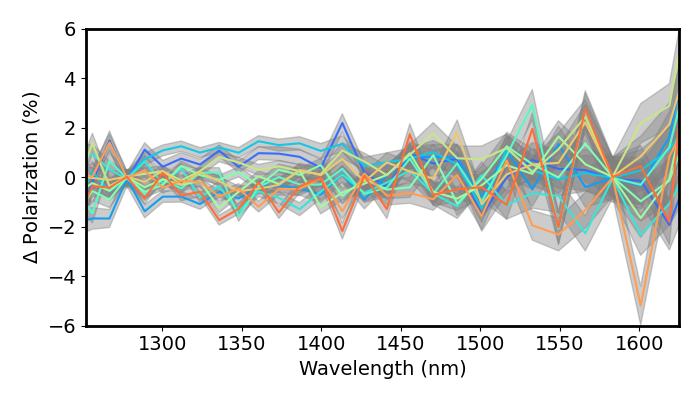}
\caption{The $\Delta$ Polarization estimated for the case of the 18th Mar 2024 Moon observation as shown in Figure \ref{fig_lunar_pol} using the same line colors. The error on the $\Delta$ polarization is shown as the grey region around the colored lines.}\label{fig_lunar_delta_pol}
\end{center}
\end{figure}

The raw spectrum (without any spectral correction) show a peak at 1300 nm, for both the detectors, which is essentially due to the spectral response of the instrument which peaks at this wavelength. A total of 13 spectra are recorded in this observation (numbered 0 to 12). The illuminated portion of the Moon and its location within the SHAPE FOV is shown in the right panel for each of the spectrum.  The spectrum in $D1$ as well as $D2$ are seen to change slightly with the changing position of the Moon in the SHAPE FOV. The observed spectra are consistent as the Moon moves in the central region of the FOV.

Next, we estimate the degree of linear polarization ($DOLP$) for this Moon observation. Without considering any effect of field dependence of $\alpha_1$ or $\alpha_2$ (i.e. assuming $\alpha_1=\alpha_2=1$), we estimate:

\begin{equation}\label{eq_DOLP}
DOLP = -\frac{1}{cos2\theta}\frac{I_1-I_2}{I_1+I_2}.
\end{equation}

Figure \ref{fig_lunar_pol} shows the $\%$ Polarization (=$DOLP$$\times$100)  estimated in this manner for all the observed spectra. This plot demonstrates the large inconsistency observed in the $DOLP$ of each observation. Firstly, for any wavelength, the $\%$ Polarization changes for each observation as the Moon moves across the SHAPE FOV and it settles to a constant value as the Moon reaches to the central region of the field. Secondly, the slope of the spectral polarization is also observed to be changing with each observation. And thirdly, the $\%$ polarization settles to a value of $\sim14\%$, which is higher than the actual lunar polarization at these phase angles ($\sim$\ang{90}) which will be in the range of $4\%$ to $7\%$ depending upon the fraction of mare and highland regions (see the discussion in Appendix \ref{appendix1_lunarpol}). Clearly, the `extra' observed polarization is the instrumental polarization which is mainly caused by field response variation across the illuminated part of the lunar disc. The spectral slopes which are observed in the in the $DOLP$ are because of $\lambda$ dependence of the field response. Despite the varying spectral slopes, the polarization spectrum is linear with wavelength and lacks any band polarization. The variations recorded in the 13 spectra as well as the polarization spectra across these observations are due to the slow drift of the SHAPE FOV, which, is shown on the right panel of the figure. As the Moon is observed to move into the center of the field, the spectral polarization is observed to be more consistent. We draw an important inference from the Moon observations that, though the absolute value of $DOLP$ changes with field, the observed spectral-polarization from the Moon can be assumed to be linearly varying with wavelength. 

In Figure \ref{fig_lunar_delta_pol}, we correct for the linear trend by taking two wavelength points at 1273 nm and 1580 nm (for all the spectra) for estimating the continuum and then subtracting the continuum to show the spectral $\Delta$ polarization over the continuum. The spectral $\Delta$ polarization of all the observations is centered at 0. As Moon lacks any molecular species in its atmosphere, we do not expect any band polarization, but it illustrates the methodology for calculating $\Delta$ polarization. The grey band around the spectrum represents the error in the estimated polarization. The method for estimating the error or the $DOLP$ uncertainty is explained in the Appendix \ref{appendix2_error}. The error on the $\Delta$ polarization is also plotted as grey region in Figure  \ref{fig_lunar_delta_pol}.

\subsection{Observations of Moon to study SHAPE field}\label{sec:Observations_of_Moon_to_study_field}

Using selected disc-integrated observations of the Moon by SHAPE, here we try to study the field response of SHAPE to estimate the variation in $\alpha_2/\alpha_1$ in the entire field. Although the laboratory study of the SHAPE field has also allowed us to study field response of SHAPE and has clearly demonstrated the non-uniformities in the SHAPE field, the laboratory measurements do not have a continuous coverage in the field and hence it is advisable to use lunar observations instead. During each SHAPE observation of the Moon, due to random drift of the satellite, the Moon has moved into various regions of the FOV. By carrying out several observations of the Moon, we have selected 5 observations here for further investigation where Moon has covered various regions of the FOV. In each of the Moon observations, by knowing the fraction of the highlands and the mare region, and knowing the phase angle of the Moon, we estimate the `actual' degree of polarization $p$ of the Moon, at the instant of observation, by the procedure explained in Appendix~\ref{appendix1_lunarpol}. For all the Moon observations presented in this paper, the fraction of highland and mare regions along with the phase angle is provided in Table \ref{table_moon_obs}. Next, we use Eqs. \ref{eq_signal1_FM} and \ref{eq_signal2_FM}, to obtain the following relation of $\alpha_2/\alpha_1$ with the incident $p$:

\begin{equation}\label{eq_DOLP_estim}
\frac{\alpha_2}{\alpha_1} = \frac{I_2(1-p)}{I_1(1+p)},
\end{equation}

where $p=-Qcos2\theta /I$ and $I_1$ and $I_2$ are the dark (background) subtracted intensities recorded in the two detectors. It is noteworthy here that since we are using the Lunar observations to study the SHAPE response of one channel with respect to another (via $\alpha_2/\alpha_1$) the individual units of $I_1$ and $I_2$ do not matter as long as the units are identical.

\begin{figure}[htbp]
    \centering
    \begin{subfigure}{\linewidth}
        \centering
        \includegraphics[width=0.9\linewidth]{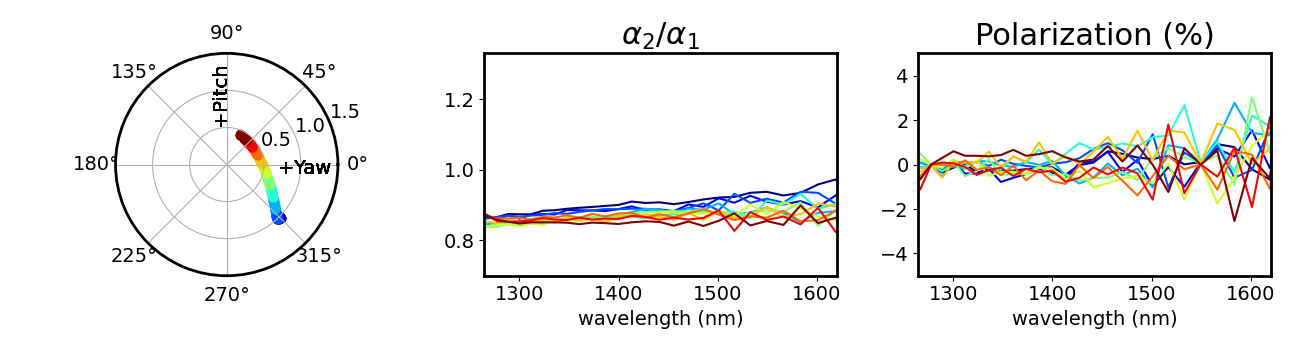}
    \end{subfigure}
    
    \begin{subfigure}{\linewidth}
        \centering
        \includegraphics[width=0.9\linewidth]{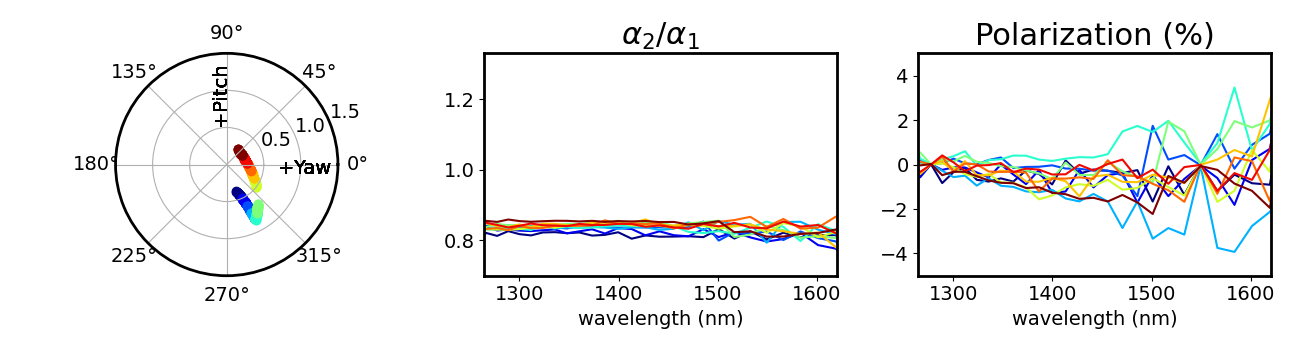}
    \end{subfigure}
    
    \begin{subfigure}{\linewidth}
        \centering
        \includegraphics[width=0.9\linewidth]{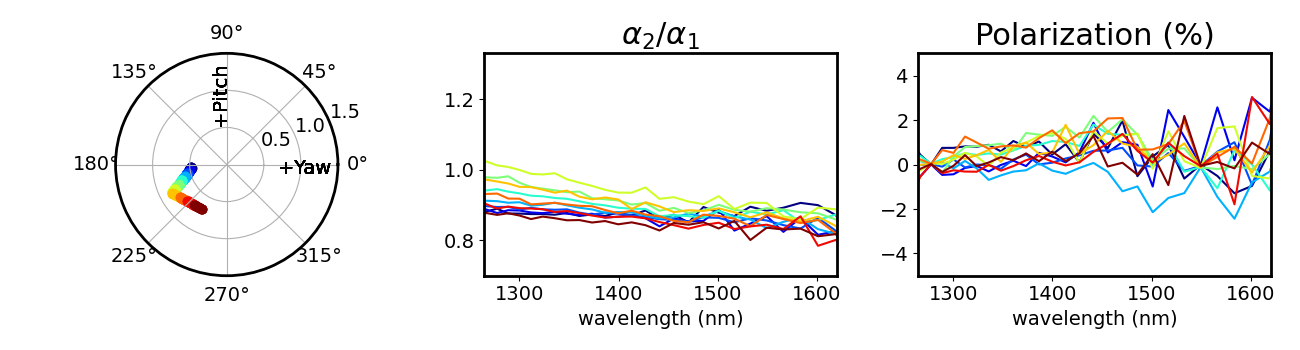}
    \end{subfigure}
    
    \begin{subfigure}{\linewidth}
        \centering
        \includegraphics[width=0.9\linewidth]{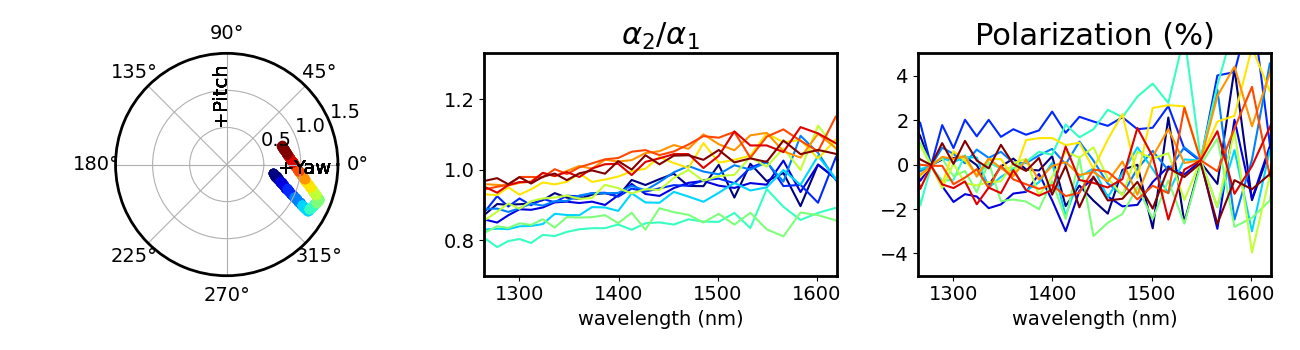}
    \end{subfigure}
    
    \begin{subfigure}{\linewidth}
        \centering
        \includegraphics[width=0.9\linewidth]{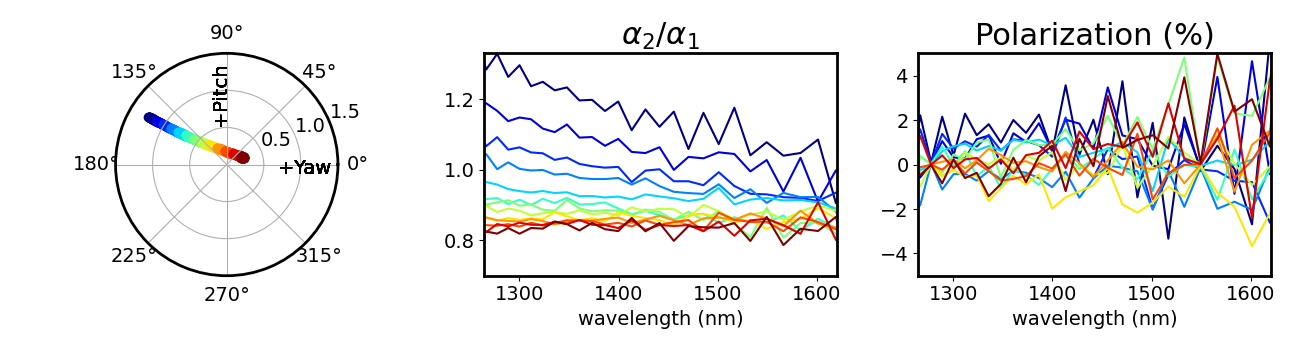}
    \end{subfigure}

    \caption{Selected Moon observations with SHAPE and estimation of $\alpha_2/\alpha_1$. The position of the Moon center in SHAPE's circular FOV is shown on the left column. The direction of spacecraft Yaw and Pitch axes are marked. As the Moon moves in different regions of the FOV the variation of $\alpha_2/\alpha_1$ is plotted in the middle column. The $\Delta$polarization for all the cases, similar to Figure \ref{fig_lunar_delta_pol}, is plotted on the right column. The color of the plot lines corresponds to the color of the Moon position plotted on the left panel. The details of all the observations are given in Table \ref{table_moon_obs} from top (No. 1) to bottom (No. 5).}
    \label{fig:Moon_SHAPE_FOV}
\end{figure}

\begin{figure}[H]
\begin{center}
\includegraphics[scale=0.8]{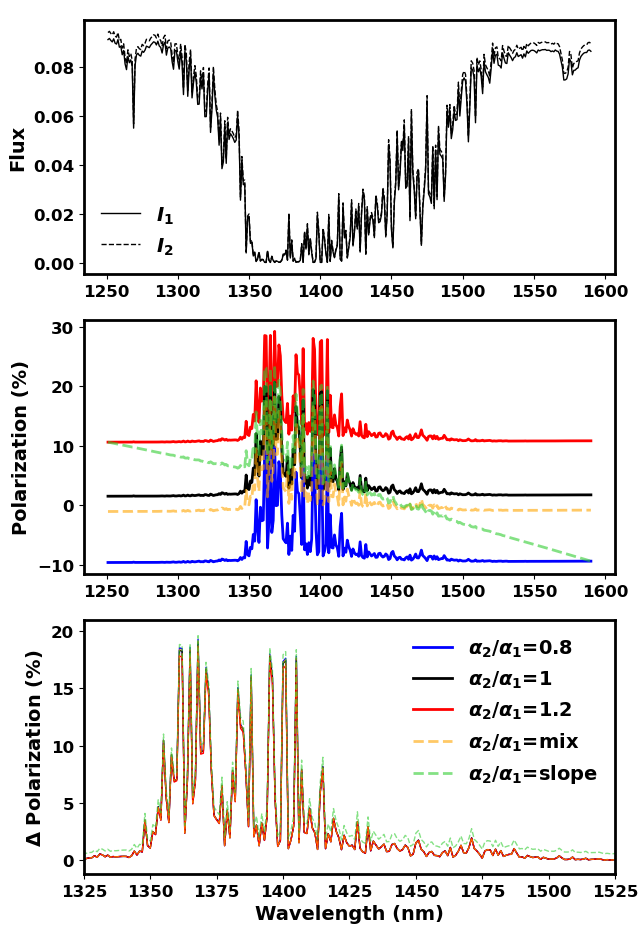}
\caption{The effect of non-identical transmission (given by $\alpha_2 / \alpha_1$) on the Earth polarization measurements. [Top] The horizontally and vertically polarized fluxes of Earth as obtained from Ref. \citenum{2022A&A...664A.172T} using Eq. \ref{eq_signal2}. [Middle] The polarization as measured by the instrument for different values of $\alpha_2 / \alpha_1$. The true polarization is given by $\alpha_2 / \alpha_1$=1 case. [Bottom] The band polarization i.e. $\Delta$ Polarization for different cases of $\alpha_2 / \alpha_1$. Refer to the text for details of each case. The middle and bottom panels share the same legend which is given in the bottom panel.} \label{fig_pol_response_spec}
\end{center}
\end{figure}

As is seen in Figure \ref{fig:Moon_SHAPE_FOV}, the $\alpha_2/\alpha_1$ lies between 0.8 to 1.2 for a large range of field positions of the Moon with the SHAPE FOV. Also, it is noteworthy that the central regions of the field have $\alpha_2/\alpha_1$ within 0.8 to 0.9, as seen in the first and second panels of Figure \ref{fig:Moon_SHAPE_FOV}. The edges of the field, however, can produce increasingly larger deviations, as demonstrate in the last panel of Figure \ref{fig:Moon_SHAPE_FOV}. As discussed in Section \ref{sec:ins_pol_model}, the variation of $\alpha_2/\alpha_1$ in the range of 0.8 to 1.2 can cause a maximum error of $10\%$ in the measured value of the $\Delta$ polarization i.e. $10\%$ \textit{of} the measured $\Delta$ polarization.

\begin{table}[h!]
\centering
\begin{tabular}{|c|c|c|c|c|}
\hline
\textbf{No.} & \textbf{Observation} & \textbf{Fraction of highland} & \textbf{Phase angle ($\phi$)} & \textbf{Lunar disc size}\\
\hline
1 & 2023-12-30 Obs-1 & 0.74 & $25.5^\circ$ & $0.69^\circ$\\
2 & 2023-12-30 Obs-2 & 0.73 & $23.9^\circ$ & $0.66^\circ$\\
3 & 2024-01-01 Obs-1 & 0.75 & $23.4^\circ$ & $0.53^\circ$\\
4 & 2024-03-18 Obs-1 & 0.72 & $100.35^\circ$ & $0.96^\circ$\\
5 & 2024-03-18 Obs-2 & 0.75 & $100.6^\circ$ & $0.94^\circ$\\
\hline
\end{tabular}
\caption{Details of the selected Moon observations which are shown in Figure \ref{fig:Moon_SHAPE_FOV}.} \label{table_moon_obs}
\end{table}

\subsection{Effect of field variation on Earth polarization measurements}\label{sec:Earth_polarization_measurement}

The clouds and ocean glints are the major sources of polarized flux reflected from Earth in the near-infrared region. The degree of polarization changes with phase angle. Apart from the continuum, the absorption bands (such as $\mathrm{H_2O}$, $\mathrm{O_2}$ and $\mathrm{CO_2}$) can have distinct polarization. This can lead to band polarization, or, as we refer to it here, $\Delta$ polarization with respect to the continuum.

Considering the non-identical transmission in the two channels of SHAPE, we attempt here to simulate its effect on the measured polarization of Earth. The polarized flux spectra of $Q$ and $I$ are obtained from Ref. \citenum{2022A&A...664A.172T} (available via FTP) for $\ang{30}$ phase angle for a fiducial case of an ocean planet with cloud covered glint having a wind velocity of 7m/s. This gives us the flux falling on the aperture of the instrument, as:

\begin{equation}\label{eq_signal2}
{I_1} =(I+Q)/2 \quad \mathrm{\&} \quad {I_2} = (I-Q)/2.
\end{equation}
The obtained $I_1$ and $I_2$ are shown in the top panel of Figure \ref{fig_pol_response_spec} for the water vapor band of 1400 nm. This flux is not yet multiplied by the transmission efficiency of the two channels of the instrument and hence the measured $DOLP$, at the instrument, can be calculated (after multiplying the transmission efficiencies) in the following manner:

\begin{equation}\label{eq_DOLP2}
DOLP = \frac{\alpha_2I_2 - \alpha_1I_1}{\alpha_2I_2 + \alpha_1I_1} \quad \mathrm{=} \quad \frac{(\alpha_2/\alpha_1)I_2 - I_1}{(\alpha_2/\alpha_1)I_2 + I_1}.
\end{equation}

Considering extreme values of $\alpha_2 / \alpha_1$, as seen from the Lunar observations, the $DOLP$ is calculated for three cases, i.e. $\alpha_2 / \alpha_1$ = 0.8, 1 and 1.2. The obtained $DOLP$ is plotted in middle panel of Figure \ref{fig_pol_response_spec}. As is evident, the continuum changes significantly with $\alpha_2 / \alpha_1$. The case of $\alpha_2 / \alpha_1$=1 is the actual $DOLP$. The lower and higher values of $\alpha_2 / \alpha_1$ reduce and increase the polarization respectively,  in agreement with the analysis presented in left panel of Figure \ref{fig_DOLP} and the overall spectrum is seen to shift up or down with $\alpha_2 / \alpha_1$. Though the absolute polarization is dependent on the value of $\alpha_2 / \alpha_1$, the band polarization with respect to the continuum polarization does not seem to be affected significantly by the $\alpha_2 / \alpha_1$. The $\Delta$ polarization i.e. band polarization with respect to the continuum polarization, is shown at the bottom panel of Figure \ref{fig_pol_response_spec} for all three cases of $\alpha_2 / \alpha_1$. The major differences in the three cases are seen at the peaks of polarization but the difference is $<1\%$. This is in agreement with the analysis presented in the right panel of Figure \ref{fig_DOLP} for the case of $\Delta P=0.2$ and $p1=0.2$ (the blue dashed curve).

We further include the effect of variation of $\alpha_2 / \alpha_1$ across the FOV of SHAPE, as seen in the Moon observations in Figure \ref{fig:Moon_SHAPE_FOV}. As Earth is an extended object, variations of  $\alpha_2 / \alpha_1$ can lead to different portions of Earth disc seeing different values of  $\alpha_2 / \alpha_1$. We consider a `mix' case where $50\%$, $25\%$ and $25\%$  of illuminated Earth disc portions are overlapping with  $\alpha_2 / \alpha_1$ values of 0.8, 1 and 1.2 respectively. The $DOLP$ for this `mix' case can then be calculated in the following manner:

\begin{equation}\label{eq_DOLP3}
DOLP = \frac{0.5(\alpha_2/\alpha_1)_1I_2 +0.25(\alpha_2/\alpha_1)_2I_2 +0.25(\alpha_2/\alpha_1)_3I_2 - I_1}{0.5(\alpha_2/\alpha_1)_1I_2 +0.25(\alpha_2/\alpha_1)_2I_2 +0.25(\alpha_2/\alpha_1)_3I_2 + I_1},
\end{equation}

where the subscripts 1, 2 and 3 on $\alpha_2 / \alpha_1$ signify the values of 0.8, 1 and 1.2 respectively. The polarization and $\Delta$ polarization are shown in the Figure \ref{fig_pol_response_spec} for the `mix' case. It is observed that the $DOLP$ is shifted up or down by the fractional weighting of the $\alpha_2 / \alpha_1$ and the effect on the $\Delta$ polarization remains $<1\%$. Next, we consider a case with spectral slope in the $\alpha_2 / \alpha_1$ which is observed in some cases of Moon observations as seen in Figure \ref{fig:Moon_SHAPE_FOV}. Although, the observed variation in $\alpha_2 / \alpha_1$ across the wavelength range are not seen to be more than 0.2 at a single location, for demonstrating the robustness of retrieving $\Delta$ polarization, we consider a higher value of 0.4 and linearly vary $\alpha_2 / \alpha_1$ from 1.2 to 0.8 across the wavelength range of 1250 to 1580 nm. The calculated $DOLP$ for this case, as expected, has a spectral slope. For calculating the $\Delta$ polarization, the spectral slope (based on two continuum points) is removed from $DOLP$ and the retrieved $\Delta$ polarization is plotted in Figure \ref{fig_pol_response_spec}. The retrieved $\Delta$ polarization, as seen, for the `slope' case shows the highest variation compared to all the previous cases considered but is still within $1\%$ of the `true' case of $\alpha_2 / \alpha_1$=1 demonstrating the robustness in the retrieval of $\Delta$ polarization in SHAPE observations.

\begin{figure}[h!]
\begin{center}
\includegraphics[trim={4cm 0cm 4cm 0cm}, clip=true, width=0.9\textwidth]{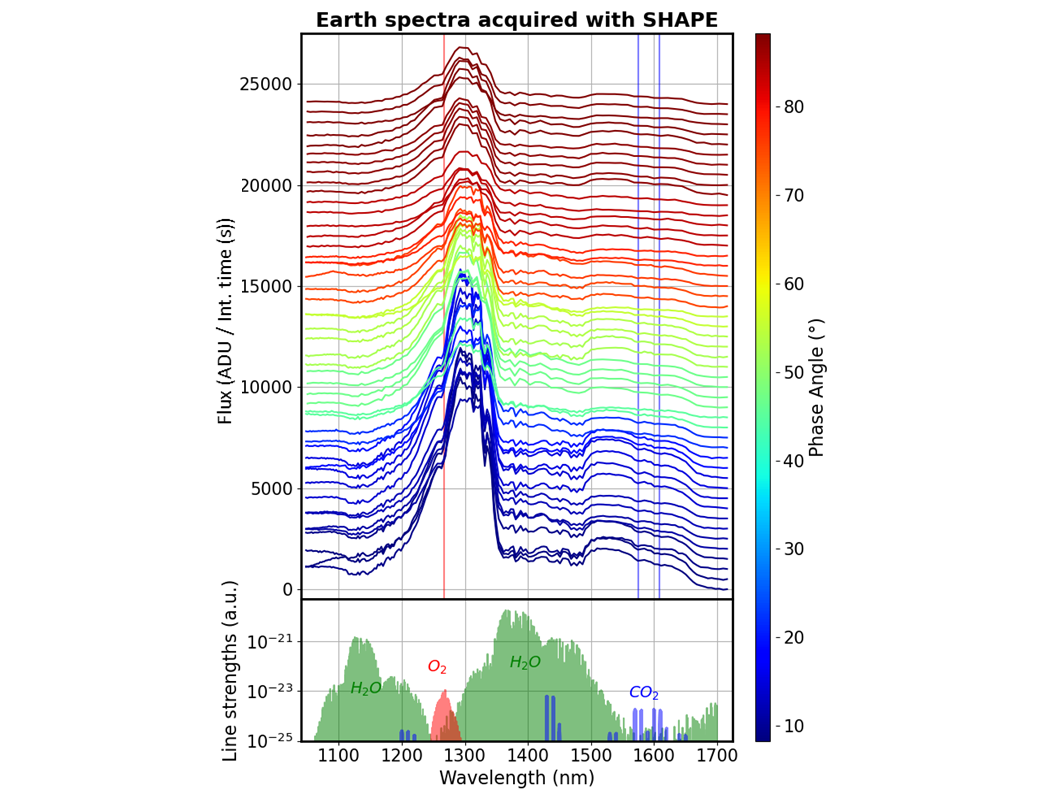}
\caption{Earth spectra, across a range of phase angles, acquired with SHAPE. The spectra are in the units of detector ADC output (ADU) divided by the integration time in seconds. Each spectrum is shifted up by 500 units for clarity. For reference, we plot the HITRAN \cite{Gordon2022HITRAN2020} line strengths of $\mathrm{H_2O}$, $\mathrm{O_2}$ and $\mathrm{CO_2}$ in arbitrary units. Without modifying the spectral locations of the lines, we shift the vertical positions of $\mathrm{O_2}$ and $\mathrm{CO_2}$ bands for clarity and show the line strengths in arbitrary units.}\label{fig_SHAPE_Earth_spectra}
\end{center}
\end{figure}

\begin{figure}[h!]
\begin{center}
\includegraphics[scale=0.6]{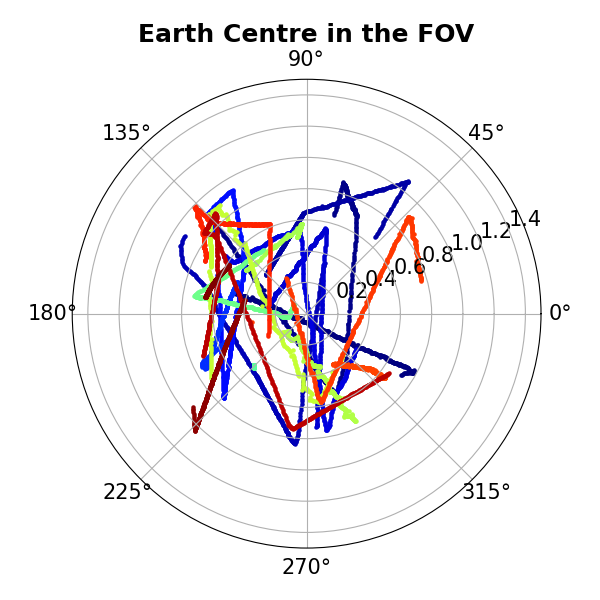}
\caption{Drift of the center of Earth within the FOV of SHAPE during the spectral scanning for the spectra which are shown in Figure \ref{fig_SHAPE_Earth_spectra}. The colors of the lines correspond to the spectra of Figure \ref{fig_SHAPE_Earth_spectra}. }\label{fig_SHAPE_Earth_drift}
\end{center}
\end{figure}

\begin{figure}[h!]
\begin{center}
\includegraphics[scale=0.6]{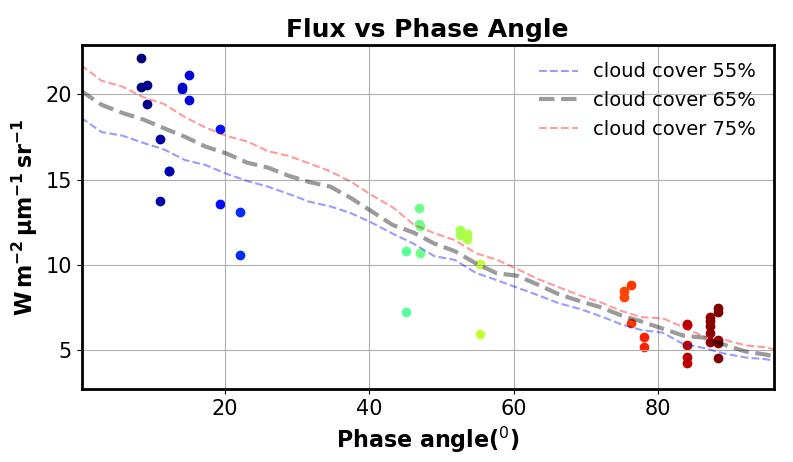}
\caption{The SHAPE measured flux of Earth at 1300 nm across phase angles as seen for the various observations shown in Figure \ref{fig_SHAPE_Earth_spectra}. The ADU values of the detector are scaled as discussed in Section \ref{sec:Earth_Obs}. Theoretically estimated relative brightness of a $55\%$, $65\%$ and $75\%$ cloudy planet, calculated using disc-integrated planet flux model of PyMieDAP \cite{2018A&A...616A.147R}, is also plotted for reference.}\label{fig_SHAPE_Earth_flux}
\end{center}
\end{figure}

\section{EARTH OBSERVATIONS WITH SHAPE}\label{sec:Earth_Obs}
Several Earth observations have been carried out with SHAPE from the orbit of Moon. As discussed above, due to non-uniform response of the field along with the random drift of Earth during the observations, the observed flux and spectra can vary within one observation and also across the observations depending upon the satellite drift. Here, we have taken a set of spectral observations in SHAPE raw data and have shown in Figure \ref{fig_SHAPE_Earth_spectra} the spectral observations of Earth across a range of phase angles. All the observations consistently show the presence of major atmospheric bands of $\mathrm{H_2O}$, $\mathrm{O_2}$ and $\mathrm{CO_2}$. The spectral continuum, as observed in all the Moon observations too, has a peak at $\sim1300$ nm and then falls down on either side of the spectrum. The two major absorption bands on either side of $\sim$ 1300 nm give all the raw spectra their characteristic large peak at $\sim1300$ nm. The spectra are plotted in the unit of detector output, that is, ADU divided by the instrument integration time (in sec). During the acquisition of these spectra the Earth has been moving in the FOV of SHAPE. The drift of the center of Earth in SHAPE field is also shown in Figure \ref{fig_SHAPE_Earth_drift} with the same colors as the spectra. In the plotted spectra no spectral or field corrections are applied here. 

Using the same observations, we plot the flux at the wavelength of 1300 nm across all the phase angles in Figure \ref{fig_SHAPE_Earth_flux}. In order to convert the detector ADC output to the unit of radiance, we use Earth observation from EPOXI mission \cite{2011AsBio..11..393R} at phase angle of \ang{57} to obtain the conversion factor of 1 ADU/sec = 0.002 $\mathrm{W/m^2/\mu m/sr}$. This conversion factor is arrived at by using the observed flux of Earth in the EPOXI mission at 1300 nm, which is 10 $\mathrm{W/m^2/\mu m/sr}$. This conversion factor is then multiplied to all the the Earth observations to convert the observations from the units of ADU/sec to $W/m^2/\mu m/sr$. Over plotted on this plot is the relative flux of a cloudy planet with different cloud fractions. As per Ref \citenum{2013BAMS...94.1031S}, the global cloud cover can vary in the range of $\sim 56\%-73\%$ and we show the variation of flux for minimum, maximum and average cloud cover. A large scatter is noticeable in the SHAPE observed flux which is mainly due to the random movement of Earth in the non-uniform field of SHAPE. However, as can also be seen in the plot, the SHAPE observations have largely captured the variation of flux with phase angles. 

This observed large scatter in the flux values can be used to study the (inconsistency in the) repeatability in the SHAPE flux measurements. As seen, the random movement of Earth in the SHAPE FOV can lead to uncertainty of $\sim 20\%$ at smaller phase angles of \ang{10} which can increase up to $\sim 50\%$ at phase angles close to \ang{90}. The variation in the flux scatter with phase angle is related to the non-uniform field response, as shown in Figure \ref{fig_field}. As seen, the non-uniformity is of \ang{1} in scale. Given the full Earth disc appears $\sim$\ang{2} in width, the larger apparent Earth, such as for smaller phase angles, will lead to larger averaging over the non-uniform portions of the field as compared to the smaller apparent Earth such as for the larger phase angles. This effect leads to variation in flux scatter with phase angles. With the help of these observations, it is possible to arrive at an empirical relation of flux scatter with phase angles assuming a linear variation in the scatter with phase angle. We find that, for the Earth observation from the Moon, this relation to be:

\begin{equation}\label{eq_Flux}
\frac{\Delta F}{F} = 0.00375\phi+0.1625, 
\end{equation}
where $F$ is the flux, $\Delta F$ denotes the scatter in the flux and $\phi$ is the phase angle in degrees. Any significant deviation observed in SHAPE fluxes, compared to the proposed $\Delta F$, may arise from physical characteristics of the Earth and warrants closer examination for its scientific significance.

\section{DISCUSSION AND CONCLUSION}\label{sec:Disc}
SHAPE, onboard Chandrayaan-3, is the first instrument to make direct measurements of Earth as an exoplanet and aims to measure not only the spectrum but also the degree of linear polarization. It uses AOTF at the heart of the instrument. The AOTFs have been previously used in other space missions too to study the planetary atmospheres. Standalone AOTF systems have been used previously in Venus and Mars missions for atmospheric studies. The SHAPE observations also demonstrate the excellent spectroscopic performance by identifying all the major gases (such as $\mathrm{H_2O}$, $\mathrm{O_2}$ and $\mathrm{CO_2}$) in the SHAPE spectral band. SHAPE instrument has been operated close to 100 times since the first observation on 21st Aug 2023 and is still functioning with the limited capability of the Chandrayaan-3 orbiter. The instrument has observed Earth from the Lunar orbit and then has been brought back to a highly elliptical Earth bound orbit which facilitates the disc-integrated observations of both Earth as well as the Moon. Several Moon observations have been carried out from the Earth bound orbit to understand the field response of SHAPE.

The laboratory characterization of the flight model of SHAPE has been studied for its spectroscopic, polarimetric as well as the field performance. We bring out and investigate the effect of the non-uniform field response of SHAPE for observing an extended object, such as Earth as seen from the Moon. We demonstrate with the help of Moon observations that the polarization measurements can suffer from an offset which can be dependent on the location of the source in the instrument field. However, all the Moon observations consistently show no band polarization i.e. the spectral polarization is observed to be flat. We establish the methodology of extracting the $\Delta$ polarization from SHAPE observations and model the offset introduced in the $\Delta$ polarization due to SHAPE. Overall, it is concluded that despite the field-dependent and pointing dependent inconsistencies, SHAPE observations can be used to study the $\Delta$ polarization in the spectra and the maximum relative error caused in the value of $\Delta$ polarization will not be more than $1\%$.

Finally, selected SHAPE observations of Earth are presented across a large range of phase angles, consistently showing the detection of absorption bands of $\mathrm{H_2O}$, $\mathrm{O_2}$ and $\mathrm{CO_2}$ in SHAPE observations. These observations demonstrate the spectroscopic performance of SHAPE and also highlight large variation in the spacecraft pointing which leads to the flux measurements having a large systematic variation in flux which can be dependent on the phase angle.

In the subsections below, we discuss more details of the Moon observations and the inference about the field response of SHAPE and also bring out the caution which need to be taken in studying the spectral polarization in Earth observations.

\subsection{Moon observations and field response}

From the Moon observations we also conclude that the non-uniformity in the field response combined with a unstable satellite pointing is the major contributor in achieving the consistency in the observations for the study of continuum flux and continuum polarization. For this reason, we refrain from presenting the SHAPE flux/spectral observations in the radiance units rather we present the observations directly in the units of the detector output (ADU). 

A study comparing the laboratory-measured field response with Lunar observations was undertaken; however, due to several limitations, the analysis remained inconclusive. An attempt was made to compare the laboratory-measured field response (measured at intervals of \ang{0.3}) with disc-integrated Lunar observations. For this purpose, a total of 19 Lunar observations obtained using SHAPE were analyzed. However, for the reasons outlined below, the comparison could not be quantified and therefore remains qualitative in nature.

Firstly, all Moon observations, like any SHAPE observations, suffer from unstable pointing, resulting in random drift of the Moon disc within the FOV. In addition, the Lunar observations were opportunistic in nature, implying that each observation had a different subtended angle, varying from $\sim\ang{0.5}$ to $\sim\ang{1.2}$, and captured different illuminated regions of the Moon. This makes inter-comparison between observations challenging, since the illuminated portion of the Moon falls on different regions of the instrument field in each case. It is also noteworthy that the brightness of the Moon varies by nearly a factor of three across the Lunar disc due to the differing albedo of the highland and mare regions making it difficult to compare two different disc-integrated observations of Moon.

Secondly, given the \ang{0.3} sampling interval used in the laboratory measurements, estimation of the laboratory-measured field response over the illuminated regions of the Moon relies heavily on interpolation between only two or three measurement points. Such interpolation is not considered reliable in view of the known non-uniformity in the field response and the strong albedo variations across the illuminated portions of the Lunar disc.

A qualitative assessment, however, allows us to understand the onboard performance of the instrument. For example, 18 March 2024 Flux observation of Moon as shown in Figure \ref{fig:Moon_SHAPE_spectrum_dark} compares well with the field variation in FOV as shown in Figure \ref{fig_field}. The Moon drift for these observations is shown in Figure \ref{fig_lunar_pol}. As Moon progresses from -Yaw (or -Azimuth) direction to +Yaw (or +Azimuth) direction, the flux is seen to increase which is in agreement with the field response as shown in Figure \ref{fig_field}. The reduction in the flux at the field edges for $D1$ as compared to $D2$ also agrees well with the field response. The values of $\alpha_2/\alpha_1$ are seen to decrease and then increase near the -Yaw (or -Azimuth) edge of the field as seen in Figure \ref{fig_field} which explains large values of $\alpha_2/\alpha_1$ as observed in the bottom panel of Figure \ref{fig:Moon_SHAPE_FOV}. It is noteworthy here that for 18 March 2024 observation, in the first three observations, small portions of the Moon, as seen in Figure \ref{fig_lunar_pol}, is outside of the $\pm{\ang{1.5}}$ circle for which the lab measurements do not exist. Given the significant size of Lunar disc in this observations, the observed $\alpha_2/\alpha_1$ values are the values averaged over the illuminated region of the Moon with contribution also from the region beyond $\pm{\ang{1.5}}$ -- where, looking at the trend, the values could be closer to 1.

\subsection{Strategy for analyzing SHAPE observations}\label{sec:SHAPE_obs_startegy}
Since SHAPE has recorded several 100s of spectra so far, we find  that out of these hundreds of spectra there are some limited observations which are consistent (i.e. having $\sim$zero drift) and will be very useful for scientific interpretation of spectral  $\Delta$ polarization of Earth (the results will be presented in an upcoming article Jaiswal et al., which is currently in preparation). Below we describe the caution which must be taken while analyzing the SHAPE observations:
\begin{itemize}
\item[1]  Repeatability of the spectrum in at least 2 consecutive spectra in both the detectors should be an essential criterion for selecting the consistent observations. Since SHAPE records the spectrum in a sequential manner (scanning from 1700 nm to 1000 nm) and not in a snapshot manner, verifying consistency across two \textit{consecutive} spectra ensures that the signal falling on the instrument has not been affected due to field variations during the observation. After ensuring the repeatability of the observations, they can be used for estimating the $\Delta$ Polarization.
\item[2]  When comparing different observations across different epochs (different orbits), especially for flux comparison, one must be careful about the location of Earth or Moon within the FOV of SHAPE. As seen in many of the Earth observations (in Figure \ref{fig_SHAPE_Earth_flux}) across different epochs and phase angles, the observed variation in flux can be significant; therefore, any comparison should properly account for systematic errors arising from the different positions of the Earth or Moon within the field of view.
\item[3] It might also be possible to estimate the absolute degree of polarization of Earth observations if the values of $\alpha_2/\alpha_1$ are known for the illuminated portion of FOV in the Earth observations from the existing Moon observations. However, it may not be a straight-forward exercise given the variation of $\alpha_2/\alpha_1$ values across FOV and the limited number of Moon observations, which may or may not exist in exactly the same region of FOV where Earth observations exist.
\end{itemize}

Overall, SHAPE has been a valuable learning experience, providing insights not only into instrument development and calibration but also into operational aspects. The SHAPE results, representing the first observations of their kind, will be presented in an upcoming research article. The lessons learned from both design and operational phases of SHAPE have been incorporated into the development of a next-generation spectro-polarimeter, VASP (Venus Atmospheric Spectro-Polarimeter) \footnote{https://www.isro.gov.in/UnionCabinetApprovesIndiasMission.html}, which builds upon and extends the legacy of SHAPE. Below are some of the specific lessons which have been incorporated to improve the design and operation of VASP: 
\begin{itemize}
    \item[1] As mentioned earlier, the observed non-uniformity in the field response is likely due to the vignetting in the optical path, mainly between the detector and the focusing lenses. The small size of the detector pixels ($500\mu m$) placed behind a narrow glass window makes focusing of the beam onto detectors a challenge. To circumvent this problem in future, a detector with larger area, i.e. 5 mm, is proposed in the optical design of VASP.
    \item[2] The significant spacecraft drift ($\sim\ang{2}$) as observed in SHAPE observations is the main reason for the observed inconsistent spectral fluxes. In future missions, the overall spacecraft pointing is expected to be significantly smaller than observed in Chandrayaan-3.
    \item[3] Due to the constraints in the mainframe system of the orbiter module, only limited onboard memory could be accommodated of $\sim$~$8$~kB per detector. This in turn resulted in limiting the duration or observations and number of spectra that could be recorded in any observation. Such memory limitations are expected to be avoided in future missions through the availability of onboard storage resources.
\end{itemize}

\appendix    % this command starts appendixes

\section{Error on DOLP}\label{appendix2_error}

In order to define the error on the DOLP, we need to propagate the errors in the observed intensities in $D1$ and D2, mentioned as $I_1$ and $I_2$. Following from Eq. \ref{eq_DOLP}, the uncertainty in DOLP, denoted \( \sigma_{\mathrm{DOLP}} \), is given by the standard error propagation formula:
\begin{equation}
    \sigma_{\mathrm{DOLP}}^2 = \left( \frac{\partial \mathrm{DOLP}}{\partial I_1} \right)^2 \sigma_{I_1}^2 + \left( \frac{\partial \mathrm{DOLP}}{\partial I_2} \right)^2 \sigma_{I_2}^2 ,
\end{equation}
where, 
\begin{align}
    \frac{\partial \mathrm{DOLP}}{\partial I_1} &= \frac{(I_1 + I_2) - (I_1 - I_2)}{(I_1 + I_2)^2} = \frac{2I_2}{(I_1 + I_2)^2} \\[10pt]
    \frac{\partial \mathrm{DOLP}}{\partial I_2} &= \frac{-(I_1 + I_2) - (I_1 - I_2)}{(I_1 + I_2)^2} = \frac{-2I_1}{(I_1 + I_2)^2}.
\end{align}
The noise in the two detectors is same and is equal to the ADC digitization noise, and hence,
\begin{equation}
    \sigma_{I_1} = \sigma_{I_2} = \sigma_I.
\end{equation}
Finally, we find that the error on the measured degree of linear polarization, \( \sigma_{\mathrm{DOLP}} \), can be written as:
\begin{equation}\label{eq_DOLP_error}
        \sigma_{\mathrm{DOLP}} = \frac{2\sigma_I \sqrt{I_1^2 + I_2^2}}{(I_1 + I_2)^2}.
\end{equation}

\begin{figure}[!htbp]
\begin{center}
\includegraphics[scale=0.6]{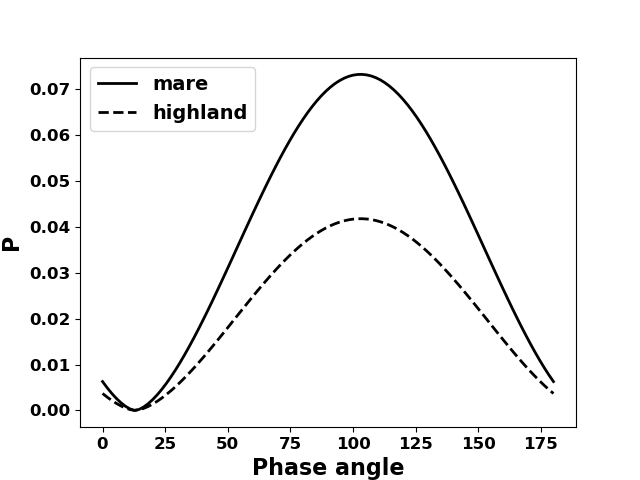}
\caption{Degree of polarization (P) of the lunar surface as a function of phase angle.}\label{fig_lunar_pol_phase_ang}
\end{center}
\end{figure}

\section{Lunar Polarization} \label{appendix1_lunarpol}
The polarization properties of planetary surfaces have been explored since several decades\cite{2024A&ARv..32....7B}. The sunlight reflected from the Moon gets polarized due to particulates present on the surface of the Moon. The degree of linear polarization, scattered by Moon, is known to have a fixed trend with phase angles and is also dependent on the albedo. The variation of polarization of a planetary surface, such as Moon, with phase angle ($\alpha$) can be approximated with a modified Rayleigh function\cite{2005SoSyR..39...45K} in the following manner:

\begin{equation}\label{eq_Lunar_pol}
P(\alpha) = \frac{[\sin^2(\alpha-\Delta\alpha)]^W}{1+\cos^2(\alpha-\Delta\alpha) + D}.
\end{equation}
Here, $W$ is the width parameter, $D$ is the depolarization parameter and the $\Delta\alpha$ is the shift parameter of the polarization maximum ($P_{max}$) from \ang{90} phase angle. This approximation is widely (such as in Ref. \citenum{2015ApJS..221...16J}) used for studying the polarization trends with phase angle. The lunar polarization measurements of the Moon do not exist in literature for the entire near-infrared wavelength range of SHAPE, and hence we rely on the measurements by Ref. \citenum{1971A&A....10...29D} who have shown the spectral dependence of the mare and highland polarization maximum ($P_{max}$) is nearly flat beyond 800 nm. For the present analysis, we assume the mare and highland polarization ($P_{max}$) to be $7\%$ and $4\%$ respectively (from measurements in Ref. \citenum{1971A&A....10...29D}) at 1050 nm wavelength in SHAPE's wavelength range of 1000-1700 nm. The value of $D$ is obtained after plugging the value of $P_{max}$, $\Delta\alpha$ (=\ang{13}\cite{2005SoSyR..39...45K}) and $W$ (=0.8\cite{2005SoSyR..39...45K}) and is found to be 22.93 and 12.65 for highland and mare regions respectively. Using this value of $D$ and $W$, one can obtain the degree of Lunar Polarization for any given phase angle and fraction of highland at the time of observation.

For all the Lunar observations discussed in this paper, the phase angle and the fraction of highland is given in Table \ref{table_moon_obs}. In our estimation of $\alpha_2/\alpha_1$ using lunar observations, we first calculate the fraction of mare and highland in the visible illuminated portion of the Moon at the time of the observation and then using Eq. \ref{eq_Lunar_pol} we find the degree of polarization contributed by the combination of mare and highlands for the observed phase angle. This estimated degree of polarization is then substituted in place of $p$ in Eq. \ref{eq_DOLP_estim} to obtain $\alpha_2/\alpha_1$.

%%%%% References %%%%%

\section*{Disclosure}
The authors declare that there are no financial interests, commercial affiliations, or other potential conflicts of interest that could have influenced the objectivity of this research or the writing of this paper.

\section*{Code and Data Availability}
The SHAPE data used in this research will be made available via https://pradan.issdc.gov.in/ch3/ after the completion of the ongoing peer review process.

\section*{Acknowledgments}
Authors thank the Deputy Director, Payload Data Mgmt. \& Space Astronomy Area; Associate Directors and Director of U. R. Rao Satellite Centre (URSC) for
encouragement and continuous support to carry out this work. Authors also thank Director, LEOS for constant support towards realization of the payload. Authors also thank the support of entire extended team of SHAPE and Chandrayaan-3, including Satyendra Kumar Singh, Kuldeep Negi, Manish Gupta, Kannan S, Muluguri Srinivasa Rao, Prapti Mittal, Sachin Narang and Himanshu Pandey. BJ thanks Gourav Mahapatra (TU, Delft, The Netherlands) for his feedback on this manuscript. Authors thank the reviewers for their comments and suggestions which have helped improve the clarity and quality of the manuscript. The Indian Space Research Organization (ISRO) funded, managed and facilitated the overall project.

\bibliography{bibliography}   % bibliography data in report.bib

@ARTICLE{2025A&A...697A.170R,
       author = {{Roccetti}, Giulia and {Emde}, Claudia and {Sterzik}, Michael F. and {Manev}, Mihail and {Seidel}, Julia V. and {Bagnulo}, Stefano},
        title = "{Planet Earth in reflected and polarized light: I. Three-dimensional radiative transfer simulations of realistic surface-atmosphere systems}",
      journal = {\aap},
         year = 2025,
        month = may,
       volume = {697},
          eid = {A170},
        pages = {A170},
          doi = {10.1051/0004-6361/202554167},
archivePrefix = {arXiv},
       eprint = {2504.02048},
 primaryClass = {astro-ph.EP},
       adsurl = {https://ui.adsabs.harvard.edu/abs/2025A&A...697A.170R}
}

@ARTICLE{1974SSRv...16..527H,
       author = {{Hansen}, J.~E. and {Travis}, L.~D.},
        title = "{Light scattering in planetary atmospheres}",
      journal = {\ssr},
         year = 1974,
        month = oct,
       volume = {16},
       number = {4},
        pages = {527-610},
          doi = {10.1007/BF00168069},
       adsurl = {https://ui.adsabs.harvard.edu/abs/1974SSRv...16..527H}
}

@ARTICLE{2022A&A...664A..21Q,
       author = {{Quanz}, S.~P. and {Ottiger}, M. and {Fontanet}, E. and {Kammerer}, J. and {Menti}, F. and {Dannert}, F. and {Gheorghe}, A. and {Absil}, O. and {Airapetian}, V.~S. and {Alei}, E. and {Allart}, R. and {Angerhausen}, D. and {Blumenthal}, S. and {Buchhave}, L.~A. and {Cabrera}, J. and {Carri{\'o}n-Gonz{\'a}lez}, {\'O}. and {Chauvin}, G. and {Danchi}, W.~C. and {Dandumont}, C. and {Defr{\'e}re}, D. and {Dorn}, C. and {Ehrenreich}, D. and {Ertel}, S. and {Fridlund}, M. and {Garc{\'\i}a Mu{\~n}oz}, A. and {Gasc{\'o}n}, C. and {Girard}, J.~H. and {Glauser}, A. and {Grenfell}, J.~L. and {Guidi}, G. and {Hagelberg}, J. and {Helled}, R. and {Ireland}, M.~J. and {Janson}, M. and {Kopparapu}, R.~K. and {Korth}, J. and {Kozakis}, T. and {Kraus}, S. and {L{\'e}ger}, A. and {Leedj{\"a}rv}, L. and {Lichtenberg}, T. and {Lillo-Box}, J. and {Linz}, H. and {Liseau}, R. and {Loicq}, J. and {Mahendra}, V. and {Malbet}, F. and {Mathew}, J. and {Mennesson}, B. and {Meyer}, M.~R. and {Mishra}, L. and {Molaverdikhani}, K. and {Noack}, L. and {Oza}, A.~V. and {Pall{\'e}}, E. and {Parviainen}, H. and {Quirrenbach}, A. and {Rauer}, H. and {Ribas}, I. and {Rice}, M. and {Romagnolo}, A. and {Rugheimer}, S. and {Schwieterman}, E.~W. and {Serabyn}, E. and {Sharma}, S. and {Stassun}, K.~G. and {Szul{\'a}gyi}, J. and {Wang}, H.~S. and {Wunderlich}, F. and {Wyatt}, M.~C. and {LIFE Collaboration}},
        title = "{Large Interferometer For Exoplanets (LIFE). I. Improved exoplanet detection yield estimates for a large mid-infrared space-interferometer mission}",
      journal = {\aap},
         year = 2022,
        month = aug,
       volume = {664},
          eid = {A21},
        pages = {A21},
          doi = {10.1051/0004-6361/202140366},
archivePrefix = {arXiv},
       eprint = {2101.07500},
 primaryClass = {astro-ph.EP},
       adsurl = {https://ui.adsabs.harvard.edu/abs/2022A&A...664A..21Q}
}

@ARTICLE{2011AsBio..11..393R,
       author = {{Robinson}, Tyler D. and {Meadows}, Victoria S. and {Crisp}, David and {Deming}, Drake and {A'Hearn}, Michael F. and {Charbonneau}, David and {Livengood}, Timothy A. and {Seager}, Sara and {Barry}, Richard K. and {Hearty}, Thomas and {Hewagama}, Tilak and {Lisse}, Carey M. and {McFadden}, Lucy A. and {Wellnitz}, Dennis D.},
        title = "{Earth as an Extrasolar Planet: Earth Model Validation Using EPOXI Earth Observations}",
      journal = {Astrobiology},
         year = 2011,
        month = jun,
       volume = {11},
       number = {5},
        pages = {393-408},
          doi = {10.1089/ast.2011.0642},
       adsurl = {https://ui.adsabs.harvard.edu/abs/2011AsBio..11..393R}
}

@ARTICLE{1993Natur.365..715S,
       author = {{Sagan}, Carl and {Thompson}, W. Reid and {Carlson}, Robert and {Gurnett}, Donald and {Hord}, Charles},
        title = "{A search for life on Earth from the Galileo spacecraft}",
      journal = {\nat},
         year = 1993,
        month = oct,
       volume = {365},
       number = {6448},
        pages = {715-721},
}

@ARTICLE{2009ApJ...700..915C,
       author = {{Cowan}, Nicolas B. and {Agol}, Eric and {Meadows}, Victoria S. and {Robinson}, Tyler and {Livengood}, Timothy A. and {Deming}, Drake and {Lisse}, Carey M. and {A'Hearn}, Michael F. and {Wellnitz}, Dennis D. and {Seager}, Sara and {Charbonneau}, David and {EPOXI Team}},
        title = "{Alien Maps of an Ocean-bearing World}",
      journal = {\apj},
         year = 2009,
        month = aug,
       volume = {700},
       number = {2},
        pages = {915-923},
archivePrefix = {arXiv},
       eprint = {0905.3742},
 primaryClass = {astro-ph.EP},
}

@ARTICLE{2018RemS...10..254Y,
       author = {{Yang}, Weidong and {Marshak}, Alexander and {V{\'a}rnai}, Tam{\'a}s and {Knyazikhin}, Yuri},
        title = "{EPIC Spectral Observations of Variability in Earth's Global Reflectance}",
      journal = {Remote Sensing},
         year = 2018,
        month = feb,
       volume = {10},
       number = {2},
        pages = {254},
          doi = {10.3390/rs10020254},
       adsurl = {https://ui.adsabs.harvard.edu/abs/2018RemS...10..254Y}
}

@ARTICLE{2003JGRD..108.4710P,
       author = {{Pall{\'e}}, E. and {Goode}, P.~R. and {Yurchyshyn}, V. and {Qiu}, J. and {Hickey}, J. and {Monta{\~n}{\'e}S Rodriguez}, P. and {Chu}, M. -C. and {Kolbe}, E. and {Brown}, C.~T. and {Koonin}, S.~E.},
        title = "{Earthshine and the Earth's albedo: 2. Observations and simulations over 3 years}",
      journal = {Journal of Geophysical Research (Atmospheres)},
         year = 2003,
        month = nov,
       volume = {108},
       number = {D22},
          eid = {4710},
        pages = {4710},
          doi = {10.1029/2003JD003611},
       adsurl = {https://ui.adsabs.harvard.edu/abs/2003JGRD..108.4710P}
}

@ARTICLE{2006ApJ...644..551T,
       author = {{Turnbull}, Margaret C. and {Traub}, Wesley A. and {Jucks}, Kenneth W. and {Woolf}, Neville J. and {Meyer}, Michael R. and {Gorlova}, Nadya and {Skrutskie}, Michael F. and {Wilson}, John C.},
        title = "{Spectrum of a Habitable World: Earthshine in the Near-Infrared}",
      journal = {\apj},
         year = 2006,
        month = jun,
       volume = {644},
       number = {1},
        pages = {551-559},
}

@ARTICLE{2012Natur.483...64S,
       author = {{Sterzik}, Michael F. and {Bagnulo}, Stefano and {Palle}, Enric},
        title = "{Biosignatures as revealed by spectropolarimetry of Earthshine}",
      journal = {\nat},
         year = 2012,
        month = mar,
       volume = {483},
       number = {7387},
        pages = {64-66},
}

@ARTICLE{2008A&A...482..989S,
       author = {{Stam}, D.~M.},
        title = "{Spectropolarimetric signatures of Earth-like extrasolar planets}",
      journal = {\aap},
         year = 2008,
        month = may,
       volume = {482},
       number = {3},
        pages = {989-1007},
          doi = {10.1051/0004-6361:20078358},
archivePrefix = {arXiv},
       eprint = {0707.3905},
 primaryClass = {astro-ph},
       adsurl = {https://ui.adsabs.harvard.edu/abs/2008A&A...482..989S}
}

@ARTICLE{2012A&A...548A..90K,
       author = {{Karalidi}, T. and {Stam}, D.~M. and {Hovenier}, J.~W.},
        title = "{Looking for the rainbow on exoplanets covered by liquid and icy water clouds}",
      journal = {\aap},
         year = 2012,
        month = dec,
       volume = {548},
          eid = {A90},
        pages = {A90},
          doi = {10.1051/0004-6361/201220245},
archivePrefix = {arXiv},
       eprint = {1211.1293},
 primaryClass = {astro-ph.EP},
       adsurl = {https://ui.adsabs.harvard.edu/abs/2012A&A...548A..90K}
}

@ARTICLE{2022A&A...664A.172T,
       author = {{Trees}, V.~J.~H. and {Stam}, D.~M.},
        title = "{Ocean signatures in the total flux and polarization spectra of Earth-like exoplanets}",
      journal = {\aap},
         year = 2022,
        month = aug,
       volume = {664},
          eid = {A172},
        pages = {A172},
          doi = {10.1051/0004-6361/202243591},
archivePrefix = {arXiv},
       eprint = {2205.05669},
 primaryClass = {astro-ph.EP},
       adsurl = {https://ui.adsabs.harvard.edu/abs/2022A&A...664A.172T}
}

@ARTICLE{2012P&SS...74..202K,
       author = {{Karalidi}, T. and {Stam}, D.~M. and {Snik}, F. and {Bagnulo}, S. and {Sparks}, W.~B. and {Keller}, C.~U.},
        title = "{Observing the Earth as an exoplanet with LOUPE, the lunar observatory for unresolved polarimetry of Earth}",
     journal = {Planetary and Space Science},
         year = 2012,
        month = dec,
       volume = {74},
       number = {1},
        pages = {202-207},
          doi = {10.1016/j.pss.2012.05.017},
archivePrefix = {arXiv},
       eprint = {1203.0209},
 primaryClass = {astro-ph.EP}
}

@ARTICLE{2021RSPTA.37990577K,
       author = {{Klind{\v{z}}i{\'c}}, D. and {Stam}, D.~M. and {Snik}, F. and {Keller}, C.~U. and {Hoeijmakers}, H.~J. and {van Dam}, D.~M. and {Willebrands}, M. and {Karalidi}, T. and {Pallichadath}, V. and {van Dijk}, C.~N. and {Esposito}, M.},
        title = "{LOUPE: observing Earth from the Moon to prepare for detecting life on Earth-like exoplanets}",
      journal = {Philosophical Transactions of the Royal Society of London Series A},
         year = 2021,
        month = jan,
       volume = {379},
       number = {2188},
          eid = {20190577},
        pages = {20190577},
          doi = {10.1098/rsta.2019.0577},
archivePrefix = {arXiv},
       eprint = {2007.16078},
 primaryClass = {astro-ph.IM},
       adsurl = {https://ui.adsabs.harvard.edu/abs/2021RSPTA.37990577K}
}

@ARTICLE{2022JATIS...8a4003B,
       author = {{Boyd}, Patricia T. and {Wilson}, Emily L. and {Smale}, Alan P. and {Supsinskas}, Pete and {Livengood}, Timothy A. and {Hewagama}, Tilak and {Villanueva}, Geronimo L. and {Marshak}, Alexander and {Krotkov}, Nickolay A. and {Pokorny}, Petr and {Bixler}, Jay and {Noland}, Jonathan D. and {Ramu}, Guru and {Cleveland}, Paul and {Ganino}, John and {Jhabvala}, Murzy and {Quintana}, Elisa and {Gilbert}, Emily and {Col{\'o}n}, Knicole and {Arney}, Giada N. and {Domagal-Goldman}, Shawn D. and {Mandell}, Avi and {Barclay}, Tom and {Kuchner}, Marc and {Ott}, Lesley},
        title = "{EarthShine: Observing our world as an exoplanet from the surface of the Moon}",
      journal = {Journal of Astronomical Telescopes, Instruments, and Systems},
         year = 2022,
        month = jan,
       volume = {8},
          eid = {014003},
        pages = {014003},
          doi = {10.1117/1.JATIS.8.1.014003},
       adsurl = {https://ui.adsabs.harvard.edu/abs/2022JATIS...8a4003B}
}

@ARTICLE{2006AsBio...6...34T,
       author = {{Tinetti}, Giovanna and {Meadows}, Victoria S. and {Crisp}, David and {Fong}, William and {Fishbein}, Evan and {Turnbull}, Margaret and {Bibring}, Jean-Pierre},
        title = "{Detectability of Planetary Characteristics in Disk-Averaged Spectra. I: The Earth Model}",
      journal = {Astrobiology},
         year = 2006,
        month = mar,
       volume = {6},
       number = {1},
        pages = {34-47},
          doi = {10.1089/ast.2006.6.34},
       adsurl = {https://ui.adsabs.harvard.edu/abs/2006AsBio...6...34T}
}

@ARTICLE{2006AsBio...6..881T,
       author = {{Tinetti}, Giovanna and {Meadows}, Victoria S. and {Crisp}, David and {Kiang}, Nancy Y. and {Kahn}, Brian H. and {Fishbein}, Evan and {Velusamy}, Thangasamy and {Turnbull}, Margaret},
        title = "{Detectability of Planetary Characteristics in Disk-Averaged Spectra II: Synthetic Spectra and Light-Curves of Earth}",
      journal = {Astrobiology},
         year = 2006,
        month = dec,
       volume = {6},
       number = {6},
        pages = {881-900},
          doi = {10.1089/ast.2006.6.881},
       adsurl = {https://ui.adsabs.harvard.edu/abs/2006AsBio...6..881T}
}

@ARTICLE{2015P&SS..113..159R,
       author = {{Rossi}, Lo{\"\i}c and {Marcq}, Emmanuel and {Montmessin}, Franck and {Fedorova}, Anna and {Stam}, Daphne and {Bertaux}, Jean-Loup and {Korablev}, Oleg},
        title = "{Preliminary study of Venus cloud layers with polarimetric data from SPICAV/VEx}",
     journal = {Planetary and Space Science},
         year = 2015,
        month = aug,
       volume = {113},
        pages = {159-168},
          doi = {10.1016/j.pss.2014.11.011}
}

@ARTICLE{2015ExA....39..445A,
       author = {{Agrawal}, Prince and {Nandi}, A. and {Sudhakar}, M. and {Jaiswal}, B. and {Tyagi}, Anurag and {Sankarasubramanian}, K. and {Agarwal}, Anil},
        title = "{Characterization of an Acousto-optic tunable filter for development of a near-IR spectrometer for planetary science}",
      journal = {Experimental Astronomy},
         year = 2015,
        month = jun,
       volume = {39},
       number = {2},
        pages = {445-460},
          doi = {10.1007/s10686-015-9461-2},
          url = {https://doi.org/10.1007/s10686-015-9461-2}
}

@BOOK{2010edpr.book.....V,
       author = {{V{\'a}zquez}, M. and {Pall{\'e}}, E. and {Rodr{\'\i}guez}, P. Monta{\~n}{\'e}s},
        title = "{The Earth as a Distant Planet}",
         year = 2010,
          doi = {10.1007/978-1-4419-1684-6},
       adsurl = {https://ui.adsabs.harvard.edu/abs/2010edpr.book.....V}
}

@ARTICLE{2013BAMS...94.1031S,
       author = {{Stubenrauch}, C.~J. and {Rossow}, W.~B. and {Kinne}, S. and {Ackerman}, S. and {Cesana}, G. and {Chepfer}, H. and {Di Girolamo}, L. and {Getzewich}, B. and {Guignard}, A. and {Heidinger}, A. and {Maddux}, B.~C. and {Menzel}, W.~P. and {Minnis}, P. and {Pearl}, C. and {Platnick}, S. and {Poulsen}, C. and {Riedi}, J. and {Sun-Mack}, S. and {Walther}, A. and {Winker}, D. and {Zeng}, S. and {Zhao}, G.},
        title = "{Assessment of Global Cloud Datasets from Satellites: Project and Database Initiated by the GEWEX Radiation Panel}",
      journal = {Bulletin of the American Meteorological Society},
         year = 2013,
        month = jul,
       volume = {94},
       number = {7},
        pages = {1031-1049},
          doi = {10.1175/BAMS-D-12-00117.1},
       adsurl = {https://ui.adsabs.harvard.edu/abs/2013BAMS...94.1031S}
}

@ARTICLE{2022JATIS...8d4007J,
       author = {{Jaiswal}, Bhavesh and {Singh}, Swapnil and {Jain}, Anand and {Sankarasubramanian}, Kasiviswanathan and {Nandi}, Anuj},
        title = "{AOTF-based spectro-polarimeter for observing Earth as an exoplanet}",
      journal = {Journal of Astronomical Telescopes, Instruments, and Systems},
         year = 2022,
        month = oct,
       volume = {8},
          eid = {044007},
        pages = {044007},
          doi = {10.1117/1.JATIS.8.4.044007},
archivePrefix = {arXiv},
       eprint = {2302.10712},
 primaryClass = {astro-ph.IM},
       adsurl = {https://ui.adsabs.harvard.edu/abs/2022JATIS...8d4007J}
}

@ARTICLE{2024A&ARv..32....7B,
       author = {{Bagnulo}, Stefano and {Belskaya}, Irina and {Cellino}, Alberto and {Kwon}, Yuna G. and {Mu{\~n}oz}, Olga and {Stam}, Daphne M.},
        title = "{Polarimetry of Solar System minor bodies and planets}",
      journal = {\aapr},
         year = 2024,
        month = dec,
       volume = {32},
       number = {1},
          eid = {7},
        pages = {7},
          doi = {10.1007/s00159-024-00157-w},
       adsurl = {https://ui.adsabs.harvard.edu/abs/2024A&ARv..32....7B}
}

@ARTICLE{2015ApJS..221...16J,
       author = {{Jeong}, Minsup and {Kim}, Sungsoo S. and {Garrick-Bethell}, Ian and {Park}, So-Myoung and {Sim}, Chae Kyung and {Jin}, Ho and {Min}, Kyoung Wook and {Choi}, Young-Jun},
        title = "{Multi-band Polarimetry of the Lunar Surface. I. Global Properties}",
      journal = {\apjs},
         year = 2015,
        month = nov,
       volume = {221},
       number = {1},
          eid = {16},
        pages = {16},
          doi = {10.1088/0067-0049/221/1/16},
       adsurl = {https://ui.adsabs.harvard.edu/abs/2015ApJS..221...16J}
}

@ARTICLE{2024arXiv241207416N,
       author = {{Nandi}, Anuj and {Singh}, Swapnil and {Jaiswal}, Bhavesh and {Jain}, Anand and {Verma}, Smrati and {Palawat}, Reenu and {Ravishankar B.}, T. and {Singh}, Brajpal and {Tyagi}, Anurag and {Das}, Priyanka and {Bose}, Supratik and {Verma}, Supriya and {Rahul Gautam}, Waghmare and {Yogesh Prasad K.}, R. and {Raha}, Bijoy and {Mendhekar}, Bhavesh and {Raju K.}, Sathyanaryana and {Rao Kondapi V.}, Srinivasa and {Kumar}, Sumit and {Thakur}, Mukund Kumar and {Bhatia}, Vinti and {Sharma}, Nidhi and {Rao Yenni}, Govinda and {Satya}, Neeraj Kumar and {Raghavendra}, Venkata and {Vivechana M.}, S. and {Leeja Justin}, Evangelin and {Karmakar}, Praloy and {Patra}, Anurag and {Manjusha J.}, Naga and {Srikanth}, Motamarri and {Rajhans}, Chinmay Kumar and {Kalpana}, K. and {P}, Veeramuthuvel},
        title = "{SHAPE -- A Spectro-Polarimeter Onboard Propulsion Module of Chandrayaan-3 Mission}",
      journal = {arXiv e-prints},
         year = 2024,
        month = dec,
          eid = {arXiv:2412.07416},
        pages = {arXiv:2412.07416},
          doi = {10.48550/arXiv.2412.07416},
archivePrefix = {arXiv},
       eprint = {2412.07416},
 primaryClass = {astro-ph.IM},
       adsurl = {https://ui.adsabs.harvard.edu/abs/2024arXiv241207416N}
}

@ARTICLE{2018A&A...616A.147R,
       author = {{Rossi}, Lo{\"\i}c and {Berzosa-Molina}, Javier and {Stam}, Daphne M.},
        title = "{PYMIEDAP: a Python-Fortran tool for computing fluxes and polarization signals of (exo)planets}",
      journal = {\aap},
         year = 2018,
        month = sep,
       volume = {616},
          eid = {A147},
        pages = {A147},
          doi = {10.1051/0004-6361/201832859},
archivePrefix = {arXiv},
       eprint = {1804.08357},
 primaryClass = {astro-ph.EP},
       adsurl = {https://ui.adsabs.harvard.edu/abs/2018A&A...616A.147R}
}

@ARTICLE{2014A&A...562L...5M,
       author = {{Miles-P{\'a}ez}, P.~A. and {Pall{\'e}}, E. and {Zapatero Osorio}, M.~R.},
        title = "{Simultaneous optical and near-infrared linear spectropolarimetry of the earthshine}",
      journal = {\aap},
         year = 2014,
        month = feb,
       volume = {562},
          eid = {L5},
        pages = {L5},
          doi = {10.1051/0004-6361/201323009},
archivePrefix = {arXiv},
       eprint = {1401.6029},
 primaryClass = {astro-ph.EP},
       adsurl = {https://ui.adsabs.harvard.edu/abs/2014A&A...562L...5M}
}

@article{ refId1,
	author = {{Emde, Claudia} and {Buras-Schnell, Robert} and {Sterzik, Michael} and {Bagnulo, Stefano}},
	title = {Influence of aerosols, clouds, and sunglint on polarization spectra of Earthshine},
	DOI= "10.1051/0004-6361/201629948",
	url= "https://doi.org/10.1051/0004-6361/201629948",
	journal = {A\&A},
	year = 2017,
	volume = 605,
	pages = "A2",
}

@article{Gordon2022HITRAN2020,
  author = {Gordon, I. E. and Rothman, L. S. and Hargreaves, R. J. and Gomez, F. M. and Bertin, T. and others},
  title = {The HITRAN2020 molecular spectroscopic database},
  journal = {Journal of Quantitative Spectroscopy and Radiative Transfer},
  volume = {277},
  pages = {107949},
  year = {2022},
  doi = {10.1016/j.jqsrt.2021.107949}
}

@ARTICLE{2025A&A...702A.262R,
       author = {{Roccetti}, Giulia and {Sterzik}, Michael F. and {Emde}, Claudia and {Manev}, Mihail and {Bagnulo}, Stefano and {Seidel}, Julia V.},
        title = "{Planet Earth in reflected and polarized light: III. Modeling and analysis of a decade-long catalog of Earthshine observations}",
      journal = {\aap},
         year = 2025,
        month = oct,
       volume = {702},
          eid = {A262},
        pages = {A262},
          doi = {10.1051/0004-6361/202555758},
archivePrefix = {arXiv},
       eprint = {2509.13415},
 primaryClass = {astro-ph.EP},
       adsurl = {https://ui.adsabs.harvard.edu/abs/2025A&A...702A.262R}
}

@ARTICLE{1968JGR....73..649B,
       author = {{Beckmann}, Petr},
        title = "{Depolarization of electromagnetic waves backscattered from the lunar surface}",
      journal = {\jgr},
         year = 1968,
        month = jan,
       volume = {73},
       number = {2},
        pages = {649-655},
          doi = {10.1029/JB073i002p00649},
       adsurl = {https://ui.adsabs.harvard.edu/abs/1968JGR....73..649B}
}

@ARTICLE{1995Natur.378..355M,
       author = {{Mayor}, Michel and {Queloz}, Didier},
        title = "{A Jupiter-mass companion to a solar-type star}",
      journal = {\nat},
         year = 1995,
        month = nov,
       volume = {378},
       number = {6555},
        pages = {355-359},
          doi = {10.1038/378355a0},
       adsurl = {https://ui.adsabs.harvard.edu/abs/1995Natur.378..355M}
}

@article{10.1117/1.JATIS.7.3.035001,
author = {Bhavesh Jaiswal},
title = {{Concept of an achromatic stellar coronagraph and its application for detecting extrasolar planets}},
volume = {7},
journal = {Journal of Astronomical Telescopes, Instruments, and Systems},
number = {3},
publisher = {SPIE},
pages = {035001},
year = {2021},
doi = {10.1117/1.JATIS.7.3.035001},
URL = {https://doi.org/10.1117/1.JATIS.7.3.035001}
}

@ARTICLE{1971A&A....10...29D,
       author = {{Dollfus}, A. and {Bowell}, E.},
        title = "{Polarimetric Properties of the Lunar Surface and its Interpretation. Part I. Telescopic Observations}",
      journal = {\aap},
         year = 1971,
        month = jan,
       volume = {10},
        pages = {29},
       adsurl = {https://ui.adsabs.harvard.edu/abs/1971A&A....10...29D}
}

@ARTICLE{2005SoSyR..39...45K,
       author = {{Korokhin}, V.~V. and {Velikodsky}, Yu. I.},
        title = "{Parameters of the positive polarization maximum of the Moon: Mapping}",
      journal = {Solar System Research},
         year = 2005,
        month = jan,
       volume = {39},
       number = {1},
        pages = {45-53},
          doi = {10.1007/s11208-005-0016-3},
       adsurl = {https://ui.adsabs.harvard.edu/abs/2005SoSyR..39...45K}
}
\bibliographystyle{spiejour}   % makes bibtex use spiejour.bst

%%%%% Biographies of authors %%%%%

%\listoffigures
%\listoftables

\end{spacing}
\end{document}